\documentclass[11pt]{article}

\usepackage{bm, commath}
\usepackage{natbib}
\usepackage{caption}
\usepackage{graphicx}
\usepackage{subcaption}
\usepackage{amsmath, amsfonts, amsthm}
\usepackage{float}
\usepackage{booktabs,siunitx}
\usepackage{url}
\usepackage{multirow}
\usepackage{amssymb}
\usepackage{bm}
\usepackage{mathrsfs}
\usepackage{xr}
\usepackage{authblk}
\usepackage{listings}

\usepackage{listings}
\allowdisplaybreaks
\usepackage{bbm}
\usepackage{textcomp}
\usepackage[margin=1in]{geometry}
\usepackage{authblk}
\usepackage[ruled]{algorithm2e}
\SetKwInput{KwParam}{Parameter}
\SetAlgoCaptionLayout{centerline}

\usepackage{sectsty}
\usepackage[onehalfspacing]{setspace}
\usepackage[usenames,dvipsnames]{color}
\usepackage[utf8]{inputenc}
\usepackage{float}
\usepackage{tikz}
\usepackage{listings}
\usetikzlibrary{intersections,decorations.pathreplacing,arrows.meta,calc,fit,patterns,shapes,shapes.misc,shapes.geometric,positioning,angles,quotes}

\allowdisplaybreaks

\newtheorem{theorem}{Theorem}

\newtheorem{assumption}{Assumption}
\newtheorem{proposition}{Proposition}

\usepackage{mathtools}

\usepackage{xcolor}

\makeatletter
\renewcommand{\algocf@captiontext}[2]{#1\algocf@typo. \AlCapFnt{}#2} 
\def\@algocf@capt@plain{top}
\renewcommand{\algocf@makecaption}[2]{%
  \addtolength{\hsize}{\algomargin}%
  \sbox\@tempboxa{\algocf@captiontext{#1}{#2}}%
  \ifdim\wd\@tempboxa >\hsize
  \hskip .5\algomargin%
  \parbox[t]{\hsize}{\algocf@captiontext{#1}{#2}}
  \else%
  \global\@minipagefalse%
  \hbox to\hsize{\box\@tempboxa}
  \fi%
  \addtolength{\hsize}{-\algomargin}%
}
\makeatother

\definecolor{dblue}{HTML}{0072B2}
\definecolor{dorange}{HTML}{D55E00}
\definecolor{dgreen}{rgb}{0.,0.6,0.}

\begin{document}

\sectionfont{\bfseries\large\sffamily}%

\subsectionfont{\bfseries\sffamily\normalsize}%

\title{
The Role of Placebo Samples in Observational Studies}

\author{Ting Ye}
\affil{Department of Biostatistics, University of Washington\thanks{tingye1@uw.edu}}

\author{Shuxiao Chen}
\affil{Department of Statistics and Data Science, University of Pennsylvania\thanks{shuxiaoc@wharton.upenn.edu}}

\author{Bo Zhang}
\affil{Vaccine and Infectious Disease Division, Fred Hutchinson Cancer Research Center\thanks{bzhang3@fredhutch.org}}

\maketitle

\begin{abstract}
In an observational study, it is common to leverage known null effect to detect bias. One such strategy is to set aside a placebo sample -- a subset of data immune from the hypothesized cause-and-effect relationship. Existence of an effect in the placebo sample raises concern of unmeasured confounding bias while absence of it corroborates the causal conclusion. This paper establishes a formal framework for using a placebo sample to detect and remove bias. We state identification assumption, and develop estimation and inference methods based on outcome regression, inverse probability weighting, and doubly-robust approaches.  Simulation studies and an empirical application illustrate the finite-sample performance of the proposed methods. 
\end{abstract}


\clearpage

\section{Introduction}
A common task of observational studies is to infer a cause-and-effect relationship. Unlike well-controlled randomized experiments where physical randomization ensures the validity of causal conclusions, observational studies are retrospective and suffer from many challenges that could compromise their causal conclusions. One major assumption researchers make is the treatment ignorability assumption (\citealp{rosenbaum1983central}), also known as the no unmeasured confounders assumption (NUCA) (\citealp{Robins1992NUCA}), which states that the exposure and control groups are comparable after adjusting for observed pre-exposure covariates. Unfortunately, concerns of bias from confounders that are not measured (e.g., lab results in the medical claims data) and cannot be measured  (e.g., motivation/personality in social science research) persist. 


How to address the unmeasured confounding bias? Broadly speaking, there are three types of strategies. First, one can conduct a sensitivity analysis that relaxes the NUCA and assesses the effect under a posited sensitivity analysis model; see, e.g., \citet{rosenbaum2002observational,vanderweele2017sensitivity}, among many others. Methods in the second category resort to a haphazard, natural experiment, one most important example being the instrumental variable (IV) methods (\citealp{AIR1996}). Methods in the third category utilize the auxiliary information about the causal mechanism and the nature of the suspected unmeasured confounder, and aim to detect, \emph{quantify} and \emph{remove} the unmeasured confounding bias. One prominent example is the methodology leveraging negative control outcomes, i.e., outcome variables not affected by the treatment (\citealp{rosenbaum1992detecting,lipsitch2010negative,Shi:2020ug}). The popular difference-in-differences (DID) estimation approach (\citealp{card1990impact}) can also be recast as a negative control outcome method \citep{sofer2016negative}. 


This article studies a method that utilizes the auxiliary information in a different way. The idea is to set aside a subset of data, which we refer to as a \emph{placebo sample}, that is immune from the hypothesized cause-and-effect relationship, analyze the exposure-outcome association in the placebo sample, and formally integrate this knowledge into estimating the causal effect. Intuitively, presence of an exposure-outcome association in the placebo sample would raise serious concerns of the NUCA, while a null association \emph{corroborates} the causal conclusion.

Placebo sample has been used as a bias detection device in empirical studies (see, e.g., \citet{hoynes2015income}) and mentioned in general discussions \citep{eggers2021placebo}. One prominent application scenario is policy evaluation, where individuals not eligible or minimally affected by the policy constitute a placebo sample. In a study of the effect of minimum wages and the earned income tax credit (EITC) on deaths of despair -- deaths due to drug overdose, suicide, and alcohol-related causes, \cite{dow2020can} use college graduates as the placebo sample because they are unlikely ``to be exposed to minimum wage jobs or to be eligible for the EITC.'' Placebo sample is a useful bias detection device; however, several limitations are evident. If we fail to detect bias, this cannot be equated with the absence of bias, as it could also be due to a lack of power, especially when the placebo sample size is small. Conversely, finding evidence of bias does not necessarily nullify the causal conclusion. The key is to formally incorporate the placebo sample in causal parameter identification, estimation and statistical inference. To the best of our knowledge, a formal statistical framework is currently lacking and this article aims to fill in this gap.






\vspace{-0.3cm}
\section{Methods}

\subsection{Nonparametric identification}
\label{subsec: nonparametric identification}
\begin{figure}
	\centering
	\begin{tabular}[b]{cc}
		\begin{tabular}[b]{c}
			\begin{subfigure}[b]{0.4\columnwidth}
				\tikzset{
					-Latex,auto,node distance =1 cm and 1 cm,semithick,
					state/.style ={ellipse, draw, minimum width = 0.7 cm},
					state1/.style ={ draw, minimum width = 0.7 cm},
					point/.style = {circle, draw, inner sep=0.04cm,fill,node contents={}},
					bidirected/.style={Latex-Latex,dashed},
					el/.style = {inner sep=2pt, align=left, sloped}
				}
				\centering
				\resizebox{!}{0.32\textwidth}{\begin{tikzpicture}
						\node[state] (a) at (0,0) {${A}$};
						\node[state] (m) [right =of a] {${M}$};
						\node[state] (y) [right =of m] {${Y}$};
						\node[state] (u) [above =of m] {$U,X$};
						\path (a) edge node[above]{} (m);
						\path (m) edge node[above]{} (y);
						\path (u) edge node[above]{} (a);
						\path (u) edge node[above]{} (y);
						\path (u) edge node[above]{} (m);
				\end{tikzpicture}} 
				\caption{DAG for the primary sample ($S=1$).}
				\label{fig:S=1}
			\end{subfigure} \\
			\begin{subfigure}[b]{0.4\columnwidth}
				\tikzset{
					-Latex,auto,node distance =1 cm and 1 cm,semithick,
					state/.style ={ellipse, draw, minimum width = 0.7 cm},
					state1/.style ={ draw, minimum width = 0.7 cm},
					point/.style = {circle, draw, inner sep=0.04cm,fill,node contents={}},
					bidirected/.style={Latex-Latex,dashed},
					el/.style = {inner sep=2pt, align=left, sloped}
				}
				\centering
				\resizebox{!}{0.32\textwidth}{\begin{tikzpicture}
						\node[state] (a) at (0,0) {${A}$};
						\node[state] (m) [right =of a] {${M}$};
						\node[state] (y) [right =of m] {${Y}$};
						\node[state] (u) [above =of m] {$U,X$};
						\path (m) edge node[above]{} (y);
						\path (u) edge node[above]{} (a);
						\path (u) edge node[above]{} (y);
						\path (u) edge node[above]{} (m);
				\end{tikzpicture}} 
				\caption{DAG for the placebo sample ($S=0$).}
				\label{fig:S=0}
			\end{subfigure}
		\end{tabular}
		&
		\begin{subfigure}[b]{0.5\columnwidth}
			\centering
			\resizebox{!}{0.53\textwidth}{
				\begin{tikzpicture}
					
					\tikzstyle{dp}=[circle, inner sep=0pt, minimum size=7pt]
					
					\tikzstyle{trobs}=[dp, draw=black, fill=black]
					\tikzstyle{trlink}=[draw=black, thick]
					
					\tikzstyle{trcf}=[dp, draw=gray, fill=gray]
					\tikzstyle{trcflink}=[draw=gray,dashed, thick]
					
					\tikzstyle{uc}=[dp, draw=dblue, fill=dblue]
					\tikzstyle{uclink}=[draw=dblue, thick]
					
					\tikzstyle{lc}=[dp, draw=dorange, fill=dorange]
					\tikzstyle{lclink}=[draw=dorange, thick]
					
					\draw [->, thick] (0,0) -- (0,4.2) {};
					\draw [->, thick] (0,0) -- (7.8,0) {}; 
					\node at (0,4.5) {$E\{Y\mid S, A, X\}$};
					\draw[thick,-] (3,-0.1) -- (3,0.1) node[anchor=north,below=5pt] {$S=0$};
					\draw[thick,-] (6,-0.1) -- (6,0.1) node[anchor=north,below=5pt] {$S=1$};
					\node[uc] (trtpre) at (3,2) {};
					\node[trobs] (trtpost) at (6,4) {};
					\node[trcf] (trtcf) at (6,3) {};
					\draw [trcflink] (trtpre) -- (trtcf);
					\draw[decoration={brace,raise=7pt},decorate,thick] (6,3) -- node[left=10pt] {Trt. eff.} (6,4);
					\node[uc] (ucpre) at (3,0.5) {};
					\node[trobs] (ucpost) at (6,1.5) {};
					\draw [uclink] (ucpre) -- (trtpre);
					\draw [trlink] (ucpost) -- (trtpost);
					\draw [trcflink] (ucpre) -- (ucpost) ;
					\node[left= of trtpre] { $A=1$};
					\node[left= of ucpre] { $A=0$};
					\draw[decoration={brace,mirror,raise=7pt},decorate,thick,dblue] (3,0.5) -- node[right=10pt] {$\Delta_0(X)$} (3,2);
					\draw[decoration={brace,mirror,raise=7pt},decorate,thick] (6,1.5) -- node[right=10pt] {$\Delta_1(X)$} (6,4);
			\end{tikzpicture}}
			\caption{\label{fig: parallal} An illustration of Assumption \ref{assump: parallel}. }
		\end{subfigure}
	\end{tabular}
	\label{fig:ABC}
	\caption{(a)-(b): A typical mechanism of placebo sample. (c) An illustration of Assumption 2. \label{fig: dag}}
\end{figure}
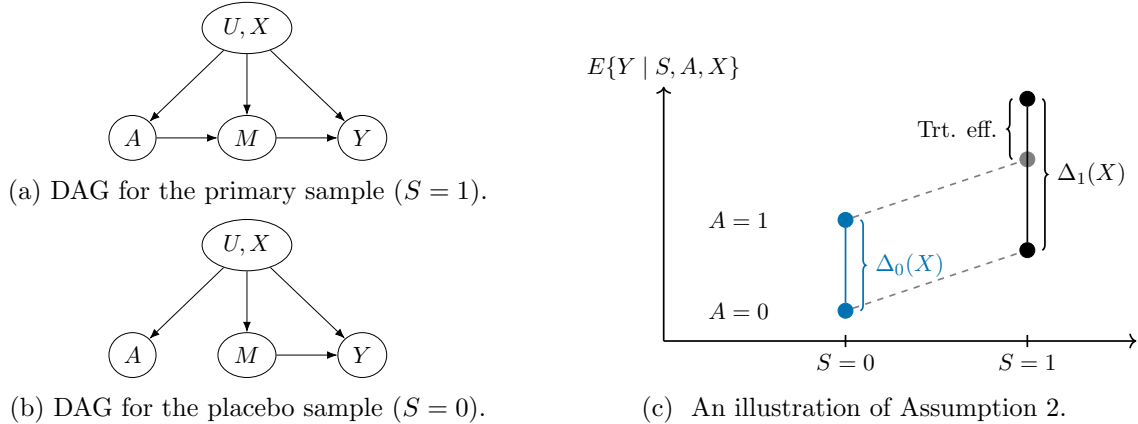

Consider a binary exposure $A \in \{0, 1\}$ and potential outcomes $\{Y^{(0)}, Y^{(1)}\}$. Throughout the article, we assume the consistency assumption and Stable Unit Treatment Value Assumption (SUTVA) so that the observed outcome $Y$ satisfies $Y= AY ^{(1)} + (1-A) Y^{(0)}$ \citep{Rubin1980}. Suppose that we have a binary baseline covariate $S$ such that $S=0$ represents the \emph{placebo sample} that is unaffected by the exposure. We call subjects with $S=1$ the \emph{primary sample}.
\begin{assumption}[Placebo sample] \label{assump: placebo samples} For $S=0$, 
	$Y^{(1)} = Y^{(0)} = Y$  almost surely.
\end{assumption}
Figure \ref{fig: dag} illustrates a typical mechanism of a placebo sample using directed acyclic graphs (DAGs), where $A$'s (exposure) effect on $Y$ (outcome) is exclusively mediated by $M$ (mediator), and $U$ encodes the unmeasured confounders of the $A$-$Y$ relationship. For individuals in the placebo sample, $A$ does not affect $M$ and thus has no effect on $Y$. For instance, in the EITC study, $A$ encodes whether an individual's state of residence has enacted EITC laws, $M$ whether she claimed the EITC, and $Y$ whether she died of death of despair. A subset of state residents, e.g., college graduates, are unlikely to be eligible for the EITC and thus cannot claim the EITC with or without EITC laws, i.e., $A$ has no effect on $M$ (as illustrated in Figure \ref{fig: dag}b).

Denote the other baseline covariates as $X$. Suppose that a random sample of $n$ subjects is obtained,  which is written as $\{ (Y_i, A_i, X_i, S_i): i=1,\dots, n\}$ and assumed to be independent and identically distributed according to the joint law of $(Y^{(A)}, A, X, S)$. Our target parameter is the average treatment effect of the treated in the primary sample
\begin{align}
	\theta_0 = E\{ Y^{(1)}- Y^{(0)} \mid S=1, A=1 \}. \label{eq: theta0}
\end{align}

Assumption \ref{assump: parallel} is our key identification assumption that links the subjects with $S=0$ and $S=1$. Figure \ref{fig: parallal} illustrates the idea. 

\begin{assumption} [Additive equi-confounding] \label{assump: parallel}
	$E\{ Y^{(0)}\mid S=1, A=1, X\} - E\{ Y^{(0)} \mid S=1, A=0, X\} = E\{ Y^{(0)} \mid S=0, A=1, X\} - E\{ Y^{(0)} \mid S=0, A=0, X\}$ almost surely.
\end{assumption}

The left-hand side of Assumption \ref{assump: parallel} encodes the level of confounding bias. Because the observed covariates $X$ may fail to render the exposure and control groups in the $S=1$ stratum comparable, $E\{ Y^{(0)}\mid S=1, A=1, X\} \neq E\{ Y^{(0)} \mid S=1, A=0, X\}$ in general. However, Assumption \ref{assump: parallel} states that the extent of residual confounding bias in the $S=1$ stratum is precisely equal to that in $S=0$, making it possible to debias using the placebo sample. 

A similar assumption also appears in the difference-in-differences (DID) and negative control outcome (NCO) literature. With repeated cross sectional data, DID assumes a parallel trends assumption which is a special case of Assumption \ref{assump: parallel} that uses the pre- and post-exposure time indicator as $S$. As such, DID with repeated cross sectional data can be recast as a placebo sample approach that uses a sample collected before the onset of the treatment as a placebo sample. For the NCO literature, the additive equi-confouding assumption says that the residual confounding bias for a NCO and the outcome is the same \citep{sofer2016negative}, compared to which Assumption \ref{assump: parallel} may be more reasonable as it is about the same outcome and is invariant to scaling of the outcome variable.  More discussion and sensitivity analysis of the assumptions are in  Supplement \S S1.  



\begin{assumption}[Positivity] \label{assump: positive}
	For some $\epsilon>0$, $P(A= 1, S=1)>\epsilon$,  $P(S=1\mid X) <1 - \epsilon$, $\epsilon<P(A=1\mid S=0,X)<1-\epsilon$, and $P(A=1\mid S=1,X)<1-\epsilon$, with probability 1. 
\end{assumption}

We give some intuition before formally stating the identification results. For $s=0,1$, $a=0,1$, and every $x$, let $\mu_Y(s, a, x) = E\{Y\mid S=s, A=a, X=x\}$, and $ \Delta_s (x)  =\mu_Y(s, 1, x)  - \mu_Y(s, 0, x)$ be observed data functions. The causal parameter of interest can be decomposed into a contrast term $C(X)$ and a bias term $B(X)$ as follows:
\begin{equation*}
	\small
	\begin{split}
		E\{ Y^{(1)} - Y^{(0)}   \mid S = 1, A = 1, X\} 
		& = \underbrace{E\{Y^{(1)} \mid S = 1, A = 1, X\} - E\{Y^{(0)} \mid S=1, A=0, X\}}_{\text{contrast}~C(X)} \\ 
		& -\underbrace{\left[E\{Y^{(0)} \mid S=1, A=1, X\} - E\{Y^{(0)} \mid S=1, A=0, X\}\right]}_{\text{bias}~B(X)}.
	\end{split}
\end{equation*}
By the consistency assumption, the contrast function $C(X)$ equals $\Delta_1(X)$ and is identified from observed data. By Assumption \ref{assump: parallel}, the bias function $B(X)$ equals $E\{Y^{(0)} \mid S=0, A=1, X\} - E\{Y^{(0)} \mid S=0, A=0, X\}$, which then equals $\Delta_0(X)$ by Assumption \ref{assump: placebo samples}. Put together, $ E\{ Y^{(1)}- Y^{(0)} \mid S=1, A=1, X \}  = \Delta_1 (X)  - \Delta_0(X) $, and is identified from observed data. Alternatively, identification can be obtained from inverse probability weighting (IPW) that avoids modelling the outcome distribution. These two identification results are stated in Proposition \ref{prop: identification}. 


\begin{proposition}\label{prop: identification}
	(a)	Under Assumptions \ref{assump: placebo samples}-\ref{assump: parallel}, $\theta_0$ defined in \eqref{eq: theta0} is identified by 
	\[
	\theta_0=  E\left\{ \Delta_1 (X)  - \Delta_0(X)  \mid S=1, A=1 \right\}. 
	\]
	(b) Suppose that Assumptions \ref{assump: positive} also holds, then 
	\begin{align*}
		\theta_0 =  \frac{1}{E\{SA\}} E\left[ \frac{S - \pi_S(X) }{  1- \pi_S(X)  } \frac{  \pi_A(X, 1)  \{ A - \pi_A(X,S)\} }{\pi_A(X,S) \{1 - \pi_A(X,S)\}}  Y  \right] ,
	\end{align*}
	where $\pi_S(X) = P(S=1\mid X) $ and $ \pi_A(X, S) = P(A=1 \mid X, S)$. 
\end{proposition}

This and all other proofs are in Supplement \S S2. There are interesting connections among the DID, NCO and our placebo sample method. Both NCO and placebo sample methods exploit known null effect: the former leverages a placebo outcome known not to be affected by the treatment, whereas the latter utilizes a placebo population known not to be affected by the treatment. We view these two methods as complementary devices that can even be used in the same study. For instance, in the study of EITC on deaths of despair, \cite{dow2020can} use both devices - a cancer outcome as an NCO and college graduates as a placebo sample. In addition, the DID with longitudinal data can be interpreted as an NCO method that uses the pre-exposure outcome as an NCO, and the DID with repeated cross sectional data can be interpreted as a placebo sample method that uses a sample drawn before the exposure in place as the placebo sample.

We note that Proposition \ref{prop: identification}(b) generalizes the IPW method proposed by \cite{Abadie:2005}, which is applicable only when $S$ is independent of all the other variables.

\subsection{Estimation and semiparametric inference}
\label{subsec:estimation}
Proposition \ref{prop: identification} suggests two estimators. The first one is a regression-based estimator 
\begin{align*}
	\hat\theta_{\rm reg}  = \frac{1}{n_{11}} \sum_{i=1}^n S_i  A_i \{ \Delta_1(X_i; \hat \beta) -  \Delta_0(X_i;  \hat \beta)\}, 
\end{align*}
where 
$n_{11} =\sum_{i=1}^n I_{(S_i=1, A_i=1)}$, $ \Delta_s(x; \hat \beta) =  \mu_Y (s,1,x; \hat\beta) -  \mu_Y (s,0,x; \hat\beta) $, $\mu_Y(s,a,x; \beta)$ is a parametric specification of $\mu_Y(s, a, x)$, and $\hat\beta$ an estimator of $\beta$. The second is an IPW estimator 
\begin{align*}
	\hat\theta_{\rm ipw}  = \frac{1}{n_{11}} \sum_{i=1}^n \frac{S_i - \pi_S(X_i; \hat\psi) }{1- \pi_S(X_i; \hat\psi)} \frac{ \pi_A(X_i, 1; \hat\alpha) \{ A_i -  \pi_A(X_i, S_i;  \hat\alpha)\} }{  \pi_A(X_i, S_i; \hat\alpha) \{1-  \pi_A (X_i, S_i; \hat\alpha)\}} Y_i ,
\end{align*}
where $\pi_S(x; \psi)$ and $\pi_A(x, s; \alpha)$ are respectively parametric models of $\pi_S(x)$ and $\pi_A(x)$, and $\hat\psi$ and $\hat \alpha$ are respectively  estimators of $\psi$ and $\alpha$. In practice, IPW-type estimators may be unstable when the (estimated) $\pi_A(\cdot)$ and $1-\pi_S(\cdot)$ are close to zero, and a common way to stabilize the weights is by normalization \citep{robins2007comment}; see the Supplement \S S1.4 for details.

The reliability of the regression-based estimator and the IPW estimator depends on correct specification of different parts of the likelihood. The consistency of the former relies on $\mu_Y(s,a,x)$ being correctly specified by $\mu_Y(s,a,x; \beta)$, whereas the latter relies on $\pi_S(x)$ and $\pi_A(x, s)$ being correctly specified by $\pi_Y(x; \psi)$ and $\pi_A(x, s; \alpha)$. In practice, when we are uncertain about which models are correctly specified, it is of interest to develop a doubly robust estimator
that is guaranteed to be consistent and deliver valid inference about $\theta_0$ provided that either $\{\mu_Y\}$ or  $\{\pi_S, \pi_A\}$, but not necessarily both, are correctly specified. The next theorem derives the efficient influence function for $\theta$ \citep{Bickel:1993book, vanderVaart:2000book} in the nonparametric model, where no restrictions are placed on the distribution of observed data  $O= (Y, A, S, X)$. Theorem \ref{theo: EIF} also provides the basis of constructing a doubly-robust estimator.

\begin{theorem}\label{theo: EIF}
	Under Assumptions \ref{assump: placebo samples}-\ref{assump: positive} and the nonparametric model, the efficient influence function for $\theta$ is 
	\begin{align*}
		& {\rm EIF} (O; \theta) =	\frac{SA}{ E\{SA\} } \{ Y -  \mu_Y(1,0,X) -  \mu_Y(0,1,X) + \mu_Y(0,0,X) - \theta\}   \\
		&  - \frac{S (1- A) \frac{\pi_A (X, 1)}{1- \pi_A(X,1)}}{E\{SA\}} \{ Y- \mu_Y(1, 0, X)\}  - \frac{ (1-S) A \frac{\pi_A (X, 1)}{\pi_A(X,0)}\frac{\pi_S(X)}{1-\pi_S(X)} }{E\{SA\}} \{ Y- \mu_Y(0, 1, X)\} \\
		&  + \frac{ (1-S) (1-A) \frac{\pi_A (X, 1)}{1- \pi_A(X,0)}\frac{\pi_S(X)}{1-\pi_S(X)} }{E\{SA\}} \{ Y- \mu_Y(0, 0, X)\}.
	\end{align*}
\end{theorem}

The efficient influence function gives an estimator $\hat\theta_{\rm dr}$ defined as the solution to $\sum_{i=1}^n \text{EIF} (O_i; \theta, \hat\eta) = 0$, where $\hat\eta= (\hat\mu_Y, \hat\pi_A, \hat\pi_S, \hat\lambda)$ denotes the collection of nuisance parameters, and $\hat\lambda = n_{11}/n$ estimates $E\{SA\}$. We prove in Supplement \S S2 that $\hat\theta_{\rm dr}$ is doubly robust. 


Next we derive the asymptotic property of $\hat\theta_{\rm dr}$. Let $\|f\|_2 =\{ \int f^2(o) dP(o)\}^{1/2}$ denote the $L_2(P)$ norm of any real-valued function $f$, and $\|f\|_2=\sum_{j=1}^\ell \|f_j\|_2$ for any collection of real-valued functions $f= (f_1,\dots, f_\ell)$, 
where $P$ denotes the distribution of $O$. Moreover, let $\eta_0 = (\mu_{Y0}, \pi_{A0}, \pi_{S0}, \lambda_0)$ denote the true values of the nuisance parameters. 

\begin{assumption} \label{assump: donsker}
	(a) $(\hat \theta_{\rm dr}, \hat\eta) \xrightarrow{P} (\theta_0, \bar\eta) $, where $\bar\eta = (\bar\mu_Y, \bar\pi_A, \bar\pi_S, \lambda_0)$ with either $\bar\mu_Y = \mu_{Y0}$ or $(\bar\pi_A, \bar\pi_S) =(\pi_{A0}, \pi_{S0})$.
	(b) For some $\epsilon>0$, $\hat\pi_S(X)<1-\epsilon, \epsilon < \hat\pi_A(0,X)<1-\epsilon,$ and $ \hat\pi_A(1,X)<1-\epsilon$ with probability 1. 
	(c) 
	For each $\theta$ in an open subset of the real line and each $\eta$ in a metric space, let $\text{EIF}(o; \theta, \eta)$ be a measurable function such that the class of functions $\{\text{EIF}(o; \theta, \eta): |\theta - \theta_0|<\epsilon, \|\eta- \bar\eta\|_2<\epsilon \}$ is Donsker for some $\epsilon>0$, and such that $E\{ \text{EIF}(O; \theta, \eta) - \text{EIF}(O; \theta_0, \bar\eta) \}^2 \rightarrow 0 $ as $(\theta, \eta) \rightarrow (\theta_0, \bar\eta)$.  
\end{assumption}

Assumption \ref{assump: donsker}(a) describes the double robustness of our proposed estimator. Assumption \ref{assump: donsker}(b) is standard for M-estimators \citep[Chapter 5.4]{vanderVaart:2000book}.

Theorem \ref{theo: asymptotics} below summarizes the doubly robust and locally efficient property of $\hat\theta_{\rm dr}$. 

\begin{theorem} \label{theo: asymptotics}
	Under Assumptions \ref{assump: placebo samples}-\ref{assump: donsker},   $\hat\theta_{\rm dr}$ satisfies
	\begin{align*}
		&  \hat\theta_{\rm dr} - \theta_0= O_P\left\{ n^{-1/2}+\big( \| \hat\pi_A - \pi_{A0} \|_2  +  \|\hat\pi_S - \pi_{S0} \|_2  \big) \| \hat\mu_Y -\mu_{Y0} \|_2 \right\},
	\end{align*}
	Suppose further that $ \big(  \| \hat\pi_A - \pi_{A0} \|_2  +  \|\hat\pi_S - \pi_{S0} \|_2  \big) \| \hat\mu_Y -\mu_{Y0} \|_2 = o_P(n^{-1/2})$, then 
	\begin{align}
		\sqrt{n} (\hat\theta_{\rm dr} - \theta_0 ) \xrightarrow{d} N\left(0, E \{ {\rm EIF} (O; \theta_0, \eta_0)^2\}\right), \label{eq: efficient}
	\end{align}
	and thus semiparametric efficient. 
\end{theorem}
The first part of Theorem \ref{theo: asymptotics} characterizes the convergence rate of $ \hat\theta_{\rm dr}$. The second part of Theorem \ref{theo: asymptotics} says that if the nuisance parameters are consistently estimated with fast rate, e.g., if they are estimated using parametric methods, then their variance contributions are negligible, and $\hat\theta_{\rm dr}$ achieves the semiparametric efficiency bound. The results in Theorems \ref{theo: EIF} and \ref{theo: asymptotics} are new, which generalize the results in \cite{sant2020doubly} to allow for $S$ being dependent on $A,X$.

When \eqref{eq: efficient} holds, a plug-in variance estimator for $\sqrt{n} \hat\theta_{\rm dr}$ can be easily constructed as $n^{-1}\sum_{i=1}^n {\rm EIF} (O_i; \hat\theta_{\rm dr}, \hat\eta)^2  $. Even if \eqref{eq: efficient}  does not hold, e.g., when only the model for $\{\mu_Y \}$ or the models for $\{ \pi_S, \pi_A\}$ are correctly specified, but all the nuisance parameters are finite-dimensional and in the form of M-estimators, $\sqrt{n} \hat\theta_{\rm dr}$ is still consistent and asymptotically normal from standard M-estimation theory \citep[Chapter 6]{newey1994_handbook}. Thus, a consistent variance estimator for  $\sqrt{n} \hat\theta_{\rm dr}$ can be constructed under the M-estimation framework; see details in the Supplement \S S3. Alternatively, the nonparametric bootstrap is commonly used in practice.


Lastly, we remark that the Donsker condition in Assumption \ref{assump: donsker} can be relaxed by using sample splitting, which enables estimating the nuisance parameters using flexible data-driven or machine learning methods \citep{chernozhukov2017double}. In particular, \eqref{eq: efficient} holds as long as the nuisance parameters are estimated at faster than $n^{-1/4}$-rates.

\vspace{-0.2cm}
\section{Simulation study}
We compare the placebo sample approach to naive methods based on NUCA, and investigate the operating characteristics of various estimators proposed in \S \ref{subsec:estimation}. We simulate the full data according to the following data-generating process with sample size $n = 1000$:

\vspace{-0.2cm}

\begin{enumerate}
	\item[(a)] $X = (X_1, X_2, X_3)$ where $X_j \sim N(0, 1)$ for $j = 1, 2, 3$.
	\item[(b)] $S$ is Bernoulli with $P(S = 1 \mid X) = \text{expit}\{-X_1 - X_2 + 3X_3 - X_2 X_3\}$,
	\item[(c)] $A$ is Bernoulli with $P(A = 1 \mid X, S) = \text{expit}\{-X_1 - X_2 + X_3 + X_2 X_3 + 0.2S + 0.5\}$.
	\item[(d)] $U$ is Bernoulli with $P(U = 1 \mid X, S, A)$ satisfying (d1) $P(U = 1 \mid X, S, A) = 0.6A + 0.2$ and (d2) $P(U = 1 \mid X, S, A) = 0.6A + 0.2\times\text{sign}\{X_1 + X_2\} + 0.2$.
	\item[(e)] $Y(0) \mid X, U, S$ is Normal with unit variance and mean satisfying (e1)  $E\{Y(0) \mid X, U, S\} = -X_1 - X_2 + 0.5X_3 + 0.5X_2X_3 + 2U + 2$ and (e2) $E\{Y(0) \mid X, U, S\} = -X_1 - X_2 - (X_3 + 0.5X_2X_3)S + 2U + 2$.
	\item[(f)] For $S=0$, $Y(1) = Y(0)$. For $S=1$, (f1) $Y(1) - Y(0) = 1$ and (f2) $Y(1) - Y(0) \sim N(1, 0.5)$. The observed outcome $Y = AY(1) + (1-A)Y(0)$. The target parameter $\theta_0=1$. 
\end{enumerate}

\vspace{-0.2cm}

The observed data are $ \{(X_i, S_i, A_i, Y_i), i=1,\dots, n\}$.   Four combinations of (d) and (e) all satisfy Assumption \ref{assump: parallel}. Our simulation study can be summarized by the factorial design below:
\begin{description}
	\item \textbf{Factor 1:} Data-generating process of $U\mid X, S, A$: (d1) and (d2);
	\item \textbf{Factor 2:} Data-generating process of $Y(0) \mid X, U, S$: (e1) and (e2);
	\item \textbf{Factor 3:} Treatment effect: (f1) and (f2);
	\item \textbf{Factor 4:} Estimator of ${\theta}_0$: (I) a naive regression estimator $\hat\theta_{\rm reg, naive}$ that regresses $Y$ on $A$ and $X$ in the $S = 1$ stratum, (II) a naive AIPW estimator $\hat\theta_{\rm dr, naive}$ based on subjects in $S = 1$, and three placebo sample estimators proposed in \S \ref{subsec:estimation}: (III) regression-based estimator $\hat\theta_{\rm reg}$, (IV) stabilized IPW estimator $\hat\theta_{\rm ipw}$, and (V) doubly-robust estimator $\hat\theta_{\rm dr}$. 
\end{description}
Factors 1-3 specify $8$ scenarios, and Factor 4 specifies estimators of $\theta_0$. We further consider three placebo sample estimators under different model misspecification. All model misspecification refers to omitting an interaction term involving $X_2 X_3$ when fitting $\mu_Y$, $\pi_S$ and $\pi_A$. Table \ref{tbl: simu n = 1000 main} summarizes the simulation results for nine estimators in 3 scenarios. The remaining 5 scenarios can be found in the Supplement \S S4. Estimators are evaluated in terms of their bias, median estimated standard error, and coverage of the $95\%$ confidence interval based on the nonparametric bootstrap using $2000$ bootstrap iterations. Both naive estimators ($\hat\theta_{\rm reg, naive}$ and $\hat\theta_{\rm dr, naive}$) are largely biased due to the unmeasured confounding. Among the three proposed estimators that leverage the placebo sample,
the regression-based estimator $\hat\theta_{\rm reg}$ has the smallest variance when $\pi_Y$ is correctly specified but becomes biased when $\pi_Y$ is misspecified. The IPW estimator $\hat\theta_{\rm ipw}$ has large finite-sample bias and poor coverage even when $\{\pi_S,\pi_A\}$ are correctly specified. The doubly robust estimator $\hat{\theta}_{\rm dr}$ is approximately unbiased in all three cases, and has smaller variance when all models are correctly specified compared to when only a subset of the models are correctly specified. We recommend using the doubly robust estimator $\hat{\theta}_{\rm dr}$ based on its robustness property and simulation results. In the Supplement \S S3, we also discuss how to construct an empirical sandwich variance estimator, and provide \textsf{R} code implementing it.

\begin{table}[ht]
	\centering
	\caption{Simulation results for three scenarios. Scenario I: d = (d1), e = (e1), f = (f1). Scenario II: d = (d2), e = (e1), f = (f1). Scenario III: d = (d1), e = (e1), f = (f2). $\theta_0 = 1$ in all three scenarios. We trimmed the 1\% tail of the most extreme values of each estimator when reporting the bias.}
	\label{tbl: simu n = 1000 main}
	\resizebox{\textwidth}{!}{
		\begin{tabular}{lcccccccccc}
			\hline
			Estimator & \begin{tabular}{c}Model \\ Spec.\end{tabular} & Bias &\begin{tabular}{c}Median \\ Est. SE\end{tabular}& \begin{tabular}{c}Cov. \\ $95\%$ CI\end{tabular}& Bias &\begin{tabular}{c}Median \\ Est. SE\end{tabular}& \begin{tabular}{c}Cov. \\ $95\%$ CI\end{tabular}& Bias &\begin{tabular}{c}Median \\ Est. SE\end{tabular}& \begin{tabular}{c}Cov. \\ $95\%$ CI\end{tabular} \\ 
			\hline
			& &\multicolumn{3}{c}{Scenario I} &\multicolumn{3}{c}{Scenario II}&\multicolumn{3}{c}{Scenario III}\\
			$\hat\theta_{\rm reg, naive}$  & & 1.20 & 0.16 & 0.00\% &1.17 & 0.16 & 0.00\% &1.19 &0.16 &0.00\%\\ 
			$\hat\theta_{\rm dr, naive}$ & & 1.21 & 0.18 & 1.70\% &1.20 & 0.16 & 0.80\% &1.19 &0.18 &1.70\%\\ 
			$\hat\theta_{\rm reg}$ & \begin{tabular}{c}$\mu_Y$ \\ correct\end{tabular} & 0.00 & 0.19 & 94.0\% &0.01 & 0.19 & 94.1\% &0.00 &0.19 &95.7\%\\ 
			$\hat\theta_{\rm reg}$ & \begin{tabular}{c}$\mu_Y$ \\ incorrect\end{tabular} &  -0.36 & 0.20 & 53.5\% &-0.35 &0.20 &56.1\% &-0.36 &0.20 &54.4\%\\ 
			$\hat\theta_{\rm ipw}$ &\begin{tabular}{c}$(\pi_S, \pi_A)$ \\ correct\end{tabular} & -0.26 & 0.61 & 87.0\% &-0.19 &0.54 &89.3\% &-0.28 &0.61 &89.4\%\\ 
			$\hat\theta_{\rm ipw}$ &\begin{tabular}{c}$(\pi_S, \pi_A)$ \\ incorrect\end{tabular}& -0.52 & 0.63 & 81.5\% &-0.44 &0.54 &83.8\% &-0.50 &0.623 &82.8\%\\ 
			$\hat\theta_{\rm dr}$ & \begin{tabular}{c}All \\ correct\end{tabular} & 0.00 & 0.47 & 92.7\% &0.02 &0.43 &93.1\% &-0.07 &0.48 &94.5\%\\ 
			$\hat\theta_{\rm dr}$ &  \begin{tabular}{c}$\mu_Y$ \\ correct\end{tabular} & -0.01 & 0.61 & 92.0\% &-0.02 &0.53 &92.6\% &-0.01 &0.61 &94.3\%\\ 
			$\hat\theta_{\rm dr}$ & \begin{tabular}{c}$(\pi_S, \pi_A)$ \\ correct\end{tabular} &0.01 & 0.50 & 90.9\% &-0.01 &0.46 &92.2\% &-0.11 &0.51 &90.9\%\\ 
			\hline
	\end{tabular}}
\end{table}

\section{Application: The effect of EITC on infant health}
Following aspects of the study by \cite{hoynes2015income}, we apply our methods to study the effect of 1993 EITC reform on low birth weight among a ``high-impact sample'' consisting of single mothers age 18 and older with a high school education or less. The treated group is second- and higher order births and the control group is the first births, because firth births were exposed to relatively small EITC credit. Our placebo sample is the single mothers who are college graduates. The accompanying data in \cite{hoynes2015income} are collapsed to cells defined by state, year, parity of birth (first, second, third, fourth or greater birth to a mother), education of mother ($<$12, 12, 13-15, $>$16), race of the mother (white, black, other), ethnicity of the mother (Hispanic, non-Hispanic, missing), and age of mother (18-24, 25-35, 35+). Since the low birth weight is a binary outcome, we recovered individual-level data from the given size and fraction of low birth weight of each cell, and ended up with 811,424 subjects in the high-impact sample and 53,131 in the placebo sample. \textcolor{black}{As discussed in \S1, the high education subgroup was unlikely to be eligible for the EITC, thus making it an appealing placebo sample. In this study, there is a concern of unmeasured confounding due to pre-exposure behaviors such as smoking and drinking, which may differ between the exposure and control groups, and lead to lower birth weight.
}

For illustration, we calculate two naive estimators based on the primary sample and effective tax year 1994 (in percentage points): the naive regression estimator $\hat\theta_{\rm reg, naive}= 1.95 (\text{SE}=0.13)$ and naive doubly-robust estimator $\hat\theta_{\rm dr, naive}= 1.74 (\text{SE}=0.12)$. Both estimators adjust for the aforementioned parity and demographic variables, and two-way interactions between the demographic variables. The standard errors are based on the nonparametric bootstrap with 100 resamples. Both naive estimators indicate a harmful effect of EITC on birth weight, which is likely biased due to unmeasured confounding. In comparison, the three proposed placebo-sample estimators are: the regression-based estimator $\hat\theta_{\rm reg} =-0.57$ (SE= 0.28), the IPW estimator $\hat\theta_{\rm ipw} =-0.38$ (SE= 1.10), and the doubly robust estimator $\hat\theta_{\rm dr} =-0.67$ (SE= 0.30). The relative performance of these three estimators is similar to that in simulation. Supplement \S S1.4 discusses Assumption \ref{assump: parallel} and presents a sensitivity analysis. Three placebo-sample estimators indicate that, for second parity or higher births among the high-impact sample, the EITC leads to reduced risk of low birth weight. Based on $\hat\theta_{\rm reg}$, the risk would be 0.57 (95\% CI: 0.03 to 1.10) percentage points lower (relative to the overall mean of 10.03 percent).

\bibliographystyle{apalike}
\bibliography{reference}

\begin{thebibliography}{}

\bibitem[Abadie, 2005]{Abadie:2005}
Abadie, A. (2005).
\newblock Semiparametric difference-in-difference estimators.
\newblock {\em Review of Economic Studies}, 75(1):1--19.

\bibitem[Angrist et~al., 1996]{AIR1996}
Angrist, J.~D., Imbens, G.~W., and Rubin, D.~B. (1996).
\newblock Identification of causal effects using instrumental variables.
\newblock {\em Journal of the American Statistical Association},
  91(434):444--455.

\bibitem[Bickel et~al., 1993]{Bickel:1993book}
Bickel, P., Klaassen, C., Ritov, Y., and Wellner, J. (1993).
\newblock {\em Efficient and Adaptive Estimation for Semiparametric Models}.
\newblock Springer.

\bibitem[Card, 1990]{card1990impact}
Card, D. (1990).
\newblock The impact of the mariel boatlift on the miami labor market.
\newblock {\em ILR Review}, 43(2):245--257.

\bibitem[Chernozhukov et~al., 2017]{chernozhukov2017double}
Chernozhukov, V., Chetverikov, D., Demirer, M., Duflo, E., Hansen, C., and
  Newey, W. (2017).
\newblock Double/debiased/neyman machine learning of treatment effects.
\newblock {\em American Economic Review}, 107(5):261--65.

\bibitem[Dow et~al., 2020]{dow2020can}
Dow, W.~H., God{\o}y, A., Lowenstein, C., and Reich, M. (2020).
\newblock Can labor market policies reduce deaths of despair?
\newblock {\em Journal of health economics}, 74:102372.

\bibitem[Eggers et~al., 2021]{eggers2021placebo}
Eggers, A.~C., Tu{\~n}{\'o}n, G., and Dafoe, A. (2021).
\newblock Placebo tests for causal inference.
\newblock {\em Working paper}.

\bibitem[Hoynes et~al., 2015]{hoynes2015income}
Hoynes, H., Miller, D., and Simon, D. (2015).
\newblock Income, the earned income tax credit, and infant health.
\newblock {\em American Economic Journal: Economic Policy}, 7(1):172--211.

\bibitem[Lipsitch et~al., 2010]{lipsitch2010negative}
Lipsitch, M., Tchetgen, E.~T., and Cohen, T. (2010).
\newblock Negative controls: a tool for detecting confounding and bias in
  observational studies.
\newblock {\em Epidemiology (Cambridge, Mass.)}, 21(3):383.

\bibitem[MacKay et~al., 2022]{mackay2022association}
MacKay, E.~J., Zhang, B., Augoustides, J.~G., Groeneveld, P.~W., and Desai,
  N.~D. (2022).
\newblock {Association of Intraoperative Transesophageal Echocardiography and
  Clinical Outcomes After Open Cardiac Valve or Proximal Aortic Surgery}.
\newblock {\em JAMA Network Open}, 5(2):e2147820--e2147820.

\bibitem[MacKay et~al., 2021]{mackay2021association}
MacKay, E.~J., Zhang, B., Heng, S., Ye, T., Neuman, M.~D., Augoustides, J.~G.,
  Feinman, J.~W., Desai, N.~D., and Groeneveld, P.~W. (2021).
\newblock Association between transesophageal echocardiography and clinical
  outcomes after coronary artery bypass graft surgery.
\newblock {\em Journal of the American Society of Echocardiography},
  34(6):571--581.

\bibitem[Metkus et~al., 2021]{metkus2021transesophageal}
Metkus, T.~S., Thibault, D., Grant, M.~C., Badhwar, V., Jacobs, J.~P., Lawton,
  J., O'Brien, S.~M., Thourani, V., Wegermann, Z.~K., Zwischenberger, B.,
  et~al. (2021).
\newblock Transesophageal echocardiography in patients undergoing coronary
  artery bypass graft surgery.
\newblock {\em Journal of the American College of Cardiology}, 78(2):112--122.

\bibitem[Newey, 1994]{Newey1994}
Newey, W.~K. (1994).
\newblock The asymptotic variance of semiparametric estimators.
\newblock {\em Econometrica}, 62(6):1349--1382.

\bibitem[Newey and McFadden, 1994]{newey1994_handbook}
Newey, W.~K. and McFadden, D. (1994).
\newblock Chapter 36 large sample estimation and hypothesis testing.
\newblock 4:2111--2245.

\bibitem[Robins et~al., 2007]{robins2007comment}
Robins, J., Sued, M., Lei-Gomez, Q., and Rotnitzky, A. (2007).
\newblock Comment: Performance of double-robust estimators when" inverse
  probability" weights are highly variable.
\newblock {\em Statistical Science}, 22(4):544--559.

\bibitem[Robins, 1992]{Robins1992NUCA}
Robins, J.~M. (1992).
\newblock { Estimation of the time-dependent accelerated failure time model in
  the presence of confounding factors}.
\newblock {\em Biometrika}, 79:321--334.

\bibitem[Rosenbaum, 1992]{rosenbaum1992detecting}
Rosenbaum, P.~R. (1992).
\newblock Detecting bias with confidence in observational studies.
\newblock {\em Biometrika}, 79(2):367--374.

\bibitem[Rosenbaum, 2002]{rosenbaum2002observational}
Rosenbaum, P.~R. (2002).
\newblock {\em Observational Studies}.
\newblock Springer.

\bibitem[Rosenbaum and Rubin, 1983]{rosenbaum1983central}
Rosenbaum, P.~R. and Rubin, D.~B. (1983).
\newblock The central role of the propensity score in observational studies for
  causal effects.
\newblock {\em Biometrika}, 70(1):41--55.

\bibitem[Rubin, 1980]{Rubin1980}
Rubin, D.~B. (1980).
\newblock {Randomization analysis of experimental data: the Fisher
  randomization test comment}.
\newblock {\em Journal of the American Statistical Association}, 75:591--593.

\bibitem[Sant’Anna and Zhao, 2020]{sant2020doubly}
Sant’Anna, P.~H. and Zhao, J. (2020).
\newblock Doubly robust difference-in-differences estimators.
\newblock {\em Journal of Econometrics}, 219(1):101--122.

\bibitem[Saul and Hudgens, 2020]{saul2020calculus}
Saul, B.~C. and Hudgens, M.~G. (2020).
\newblock The calculus of m-estimation in r with geex.
\newblock {\em Journal of statistical software}, 92(2).

\bibitem[Shi et~al., 2020]{Shi:2020ug}
Shi, X., Miao, W., Nelson, J.~C., and Tchetgen~Tchetgen, E.~J. (2020).
\newblock Multiply robust causal inference with double-negative control
  adjustment for categorical unmeasured confounding.
\newblock {\em Journal of the Royal Statistical Society: Series B (Statistical
  Methodology)}, 82(2):521--540.

\bibitem[Sofer et~al., 2016]{sofer2016negative}
Sofer, T., Richardson, D.~B., Colicino, E., Schwartz, J., and Tchetgen, E.
  J.~T. (2016).
\newblock On negative outcome control of unobserved confounding as a
  generalization of difference-in-differences.
\newblock {\em Statistical science: a review journal of the Institute of
  Mathematical Statistics}, 31(3):348.

\bibitem[Stefanski and Boos, 2002]{stefanski2002calculus}
Stefanski, L.~A. and Boos, D.~D. (2002).
\newblock The calculus of m-estimation.
\newblock {\em The American Statistician}, 56(1):29--38.

\bibitem[van~der Vaart, 2000]{vanderVaart:2000book}
van~der Vaart, A. (2000).
\newblock {\em Asymptotic Statistics}.
\newblock Cambridge University Press.

\bibitem[VanderWeele and Ding, 2017]{vanderweele2017sensitivity}
VanderWeele, T.~J. and Ding, P. (2017).
\newblock Sensitivity analysis in observational research: introducing the
  e-value.
\newblock {\em Annals of internal medicine}, 167(4):268--274.

\bibitem[Zhao et~al., 2019]{zhao2019sensitivity}
Zhao, Q., Small, D.~S., and Bhattacharya, B.~B. (2019).
\newblock Sensitivity analysis for inverse probability weighting estimators via
  the percentile bootstrap.
\newblock {\em Journal of the Royal Statistical Society: Series B (Statistical
  Methodology)}, 81(4):735--761.

\end{thebibliography}

\newpage 
\begin{center}
	{\sffamily\bfseries\LARGE
Supplement to ``The Role of Placebo Samples in Observational Studies''
	}
\end{center}

	Section \ref{sec: additional detail} contains further discussion of Assumption \ref{assump: parallel}, two proposed sensitive analysis approaches, and a formula of the stabilized estimators. Section \ref{sec: proof} contains technical proofs. Section \ref{sec: variance} contains additional details on the variance estimation.

\setcounter{equation}{0}
\setcounter{table}{0}
\setcounter{lemma}{0}
\setcounter{section}{0}
\setcounter{figure}{0}
\setcounter{theorem}{0}
\setcounter{assumption}{0}
\renewcommand{\theequation}{S\arabic{equation}}
\renewcommand{\thelemma}{S\arabic{lemma}}
\renewcommand{\thetheorem}{S\arabic{theorem}}
\renewcommand{\thefigure}{S\arabic{figure}}
\renewcommand{\thesection}{S\arabic{section}}
\renewcommand{\theassumption}{S\arabic{assumption}}


\setcounter{equation}{0}
\setcounter{table}{0}
\setcounter{lemma}{0}
\setcounter{section}{0}
\setcounter{figure}{0}
\setcounter{theorem}{0}
\setcounter{assumption}{0}
\renewcommand{\theequation}{S\arabic{equation}}
\renewcommand{\thelemma}{S\arabic{lemma}}
\renewcommand{\thetheorem}{S\arabic{theorem}}
\renewcommand{\thefigure}{S\arabic{figure}}
\renewcommand{\thesection}{S\arabic{section}}
\renewcommand{\theassumption}{S\arabic{assumption}}

\setcounter{page}{1}

\section{\textcolor{black}{Additional detail}} \label{sec: additional detail}

\subsection{Further discussion of Assumption \ref{assump: parallel}} \label{sec: assumption 2}

Let $U$ denote the set of unmeasured confounders such that $A\perp (Y^{(0)}, Y^{(1)} ) \mid U, X,S$. One set of sufficient (but not necessary) conditions for Assumption \ref{assump: parallel} to hold is (i) $P(U\mid S=1, A, X) = P(U\mid S=0, A, X) $ as depicted in Figure \ref{fig:S=0}, and (ii) $E\{Y^{(0)} \mid U, X, S=1\} - E\{Y^{(0)} \mid U, X, S=0\} $ is a function of $X$ alone. In words, Assumption \ref{assump: parallel} holds when the distribution of $U$ is the same for $S = 0$ and $S=1$ within each level of $(A, X)$, and there is no additive $S$-$U$ interaction in $E\{Y^{(0)} \mid U, X, S\}$.

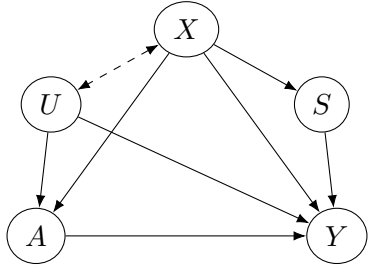
\begin{figure}[H]
	\centering
	\tikzset{
		-Latex,auto,node distance =1 cm and 1 cm,semithick,
		state/.style ={ellipse, draw, minimum width = 0.7 cm},
		state1/.style ={ draw, minimum width = 0.7 cm},
		point/.style = {circle, draw, inner sep=0.04cm,fill,node contents={}},
		bidirected/.style={Latex-Latex,dashed},
		el/.style = {inner sep=2pt, align=left, sloped}
	}
	\centering
	\resizebox{!}{!}{\begin{tikzpicture}
			\node[state] (a) at (0,0) {${A}$};
			\node[state] (y) [right =of a, xshift=2.2cm] {${Y}$};
			\node[state] (u) [above =of a, xshift=0.2cm] {$U$};
			\node[state] (s) [above =of y, xshift=-0.2cm] {$S$};
			\node[state] (x) [above =of a, xshift=2cm, yshift=1cm] {$X$};
			\path (a) edge node[above]{} (y);
			\path (u) edge node[above]{} (a);
			\path (u) edge node[above]{} (y);
			\path[bidirected] (x) edge node[above]{} (u);
			\path (x) edge node[above]{} (s);
			\path (x) edge node[above]{} (a);
			\path (x) edge node[above]{} (y);
			\path (s) edge node[above]{} (y);
	\end{tikzpicture}} 
	\caption{A DAG of a sufficient condition of Assumption  \ref{assump: parallel}}
	\label{fig:S=0}
\end{figure}

We expect Assumption \ref{assump: parallel} to hold approximately in various applications. To give an example, in a study of the effect of intraoperative TEE on patients' post-surgery clinical outcomes, there is a concern of unmeasured confounding due to surgeons' experiences as a surgeon' preference for using intraoperative TEE may depend on her experience, and a surgeon's experience may affect complication management during the surgery and hence the clinical outcome (\citealp{mackay2021association,mackay2022association}). Unfortunately, the disease registry data does not contain surgeon-level characteristics. The proposed placebo samples approach could help in this case. The healthiest cardiac surgery group, those under the age of $40$ with an ejection fraction $> 55\%$ (normal range: $55$ - $75\%$) are likely not to benefit (or benefit minimally) from the intraoperative TEE (see, e.g., empirical results from \cite{metkus2021transesophageal}). Assumption \ref{assump: placebo samples} is likely to hold for this placebo sample. On the other hand, Within the same hospital, it is largely random which surgeon operates on which patient, so that the unmeasured confounder $U$ is likely to have similar distribution for patients in the placebo sample and other patients and Assumption \ref{assump: parallel} approximately holds. Note that Assumption \ref{assump: parallel} will not likely to hold without conditioning on the hospital indicator $X = \text{hospital indicator}$. Community hospitals tend to refer complex surgeries on higher-risk patients to university hospitals; therefore, it is likely that more experienced surgeons in university hospitals (compared to those in community hospitals) would operate on higher-risk patients; however, conditioning on the hospital would render $U$ (surgeons' experience) independent of $S$ (a patient being in the placebo sample). Of course, even though both Assumption \ref{assump: placebo samples} and \ref{assump: parallel} are likely to approximately hold in this case, we cannot completely rule out the possibility of minor violation, and a sensitivity analysis could be helpful. In the following, we propose sensitivity analysis based on two different sensitivity models that may be useful in a broad range of applications. 

\subsection{Sensitivity analysis of Assumptions \ref{assump: placebo samples}-\ref{assump: parallel} under linear sensitivity models} \label{subsec: linear SA}

We propose a sensitivity analysis method of Assumptions \ref{assump: placebo samples}-\ref{assump: parallel} under linear sensitivity models. In Section \ref{subsec: nonparametric identification}, we decompose the causal parameter of interest as $\theta_0=E\{ C(X) - B(X) \mid S=1, A=1\} $, where $C(X)= \Delta_1(X)$ is identifiable from observed data. The key is in bounding $B(X)$ when Assumptions \ref{assump: placebo samples}-\ref{assump: parallel} are violated. Consider the following sensitivity model.
\begin{assumption} \label{assump: S1}
	$E\{ Y^{(1)} -Y^{(0)} \mid S=0, A=1, X \}= \lambda_0+ \lambda_X^T X$ almost surely.
\end{assumption}

\begin{assumption}\label{assump: S2}
	$E\{ Y^{(0)}\mid S=1, A=1, X\} - E\{ Y^{(0)} \mid S=1, A=0, X\} - [E\{ Y^{(0)} \mid S=0, A=1, X\} - E\{ Y^{(0)} \mid S=0, A=0, X\}]= \delta_0+ \delta_X^T X$ almost surely.
\end{assumption}
Here, $H: = \{\lambda_0, \lambda_X, \delta_0,\delta_X\} $ will be used as sensitivity parameters. When all these parameters are equal to zero, Assumptions \ref{assump: S1}-\ref{assump: S2} degenerate to Assumptions \ref{assump: placebo samples}-\ref{assump: parallel} that are imposed for the primary analysis. 

Under Assumptions \ref{assump: S1} - \ref{assump: S2}, we have that $\theta_0 = \theta_0 (H)$, where 
\begin{align*}
	\theta_0 (H) & = E\{ \Delta_1(X) - \Delta_0 (X)\mid S=1, A=1\}  \\
	&\quad -  (\delta_0 - \lambda_0 ) - (\delta_X - \lambda_X)^T E\{ X\mid S=1, A=1\}.  
\end{align*}
Fix parameters $\Gamma_L, \Gamma_U, \Lambda_L, \Lambda_U$ and consider the set of sensitivity parameters  $$ \mathcal{H} =\{ H: \delta_0, \delta_{X1}, \dots, \delta_{Xp} \in [\Gamma_{L}, \Gamma_{U}] ~ \text{and} ~ \lambda_0, \lambda_{X1},\dots, \lambda_{Xp} \in [\Lambda_{L}, \Lambda_{U}] \},$$  where $p=\text{dim}(X)$. Then $\theta_0 (H)$ is in the set 
\begin{align*}
	\Theta_0 :=   [\inf_{H\in \mathcal{H}} \theta_0 (H), \sup_{H\in \mathcal{H}}  \theta_0 (H) ].
\end{align*}
The set $\Theta_0$ is often referred to as the partially identified region under the sensitivity model depicted by Assumptions \ref{assump: S1}-\ref{assump: S2} and parameters $\Gamma_L, \Gamma_U, \Lambda_L, \Lambda_U$.

A general method to construct confidence interval for the set $\Theta_0$ is the union method. Taking an estimator $\hat \theta \in \{\hat\theta_{\rm reg},\hat\theta_{\rm ipw}, \hat\theta_{\rm dr}\}$ of  $E\{ \Delta_1(X) - \Delta_0 (X)\mid S=1, A=1\} $, an estimator of $\theta_0(H)$ is 
\begin{align}
	\hat\theta(H) & = \hat\theta - (\delta_0 -\lambda_0 ) - (\delta_X  - \lambda_X)^T \frac{1}{n_{11}} \sum_{i: S_i = A_i = 1} X_i  \nonumber \\
	&= \hat\theta - (\delta_0 -\lambda_0 ) - \sum_{j=1}^p (\delta_{Xj}  - \lambda_{Xj}) \frac{1}{n_{11}} \sum_{i: S_i = A_i = 1} X_{ij} .  \nonumber
\end{align}
Suppose that for any fixed $H$, we construct a confidence interval $[L(H), U(H)]$ for $\theta_0 (H)$, e.g., by nonparametric bootstrap, the interval $[\inf_{H\in \mathcal{H}} L(H), \sup_{H\in \mathcal{H}} U(H)] $ is an asymptotic confidence interval of $\theta_0$ with at least $(1-\alpha)$-coverage under the collection of sensitivity models $\mathcal{H}$. However, as discussed in \citep{zhao2019sensitivity}, $L(H)$ and $U(H)$ may be complicated functions of $H$, and numerical optimization over $H$ may be unstable. Therefore,
we use 
\begin{align}
	\left[ Q_{\alpha/2} ( \inf_{H\in \mathcal{H}} \hat\theta^*(H) ), Q_{1-\alpha/2} (  \sup_{H\in \mathcal{H}} \hat\theta^*(H) )\right] 
	\label{eq: SA CI}
\end{align}
as our confidence interval, where $ \inf_{H\in \mathcal{H}} \hat\theta^*(H)$ is the analogue of $\inf_{H\in \mathcal{H}} \hat\theta(H)$ computed using each bootstrap sample, and $Q_{\alpha/2} (\inf_{H\in \mathcal{H}} \hat\theta^*(H))$ is the $\alpha/2$-percentile of $ \inf_{H\in \mathcal{H}} \hat\theta^*(H)$ in the bootstrap distribution; the quantities in the upper bound are defined similarly. The interval in \eqref{eq: SA CI} is shown by \cite{zhao2019sensitivity} to cover $\theta_0$ with probability at least $1-\alpha$ asymptotically. Note that as $\hat\theta^*(H)$ is monotone in every element in $H$, the confidence interval \eqref{eq: SA CI} is computationally very simple. 

Other linear sensitivity models can also be similarly considered if having domain knowledge. For example, the considered set of sensitivity parameters $\mathcal{H}$ can be refined by imposing parameter-specific bounds if having domain knowledge. Another model that may be useful is to consider the left hand-side of Assumption \ref{assump: S2} to be positive almost surely, or negative almost surely.

\subsection{Sensitivity analysis of Assumptions \ref{assump: placebo samples}-\ref{assump: parallel} under marginal sensitivity models}

Consider the following sensitivity model. With slight abuse of notations, we still denote  the sensitivity parameters as $\Lambda$ and $\Gamma$, though they have distinct meanings compared to those in Section \ref{subsec: linear SA}. 
\begin{assumption} \label{assump: S3}
	$- \Lambda \leq E\{ Y^{(1)} -Y^{(0)} \mid S=0, A=1, X \}\leq  \Lambda$ almost surely.
\end{assumption}

\begin{assumption}\label{assump: S4}
	Let $U$ denote the set of unmeasured confounders such that $A\perp (Y^{(0)}, Y^{(1)} ) \mid U, X,S$. Assume (i)
	\begin{align}
		\frac{1}{\Gamma} \leq  \text{OR} \{\pi_S (U, A, X) , \pi_S (A, X)  \} \leq \Gamma ,\label{eq: marginal}
	\end{align}
	where $\pi_S (U, A, X) = P(S=1\mid U,A,X)$, $\pi_S (A, X) = P(S=1\mid A,X)$, and $\text{OR} (p_1, p_2) = \frac{p_1/(1-p_1)}{p_2/(1-p_2)}$;
	and (ii) $E\{Y^{(0)} \mid U, X, S=1\} - E\{Y^{(0)} \mid U, X, S=0\} $ is a function of $X$ alone.  
\end{assumption}
The sensitivity model in Assumptions \ref{assump: S3}-\ref{assump: S4} relaxes the two most disputable assumptions, and allow for the treatment to have a small effect on the placebo sample and that the distribution of $U$ being dependent on $S$ after conditioning on  $A,X$. In particular, the model in \eqref{eq: marginal} is termed the marginal sensitivity model in the literature \citep{zhao2019sensitivity}. When $\Gamma=1$, Assumption \ref{assump: S4} returns to the case where $U\perp S\mid A, X$. 

We also suppose that $E\{ Y^{(0)} \mid S=0, A, X \} \geq 0$ almost surely, which is not a restriction because we can always add a constant to every $Y_i$ to make $Y_i\geq 0$ without affecting the parameter of interest.

Let $r(A, U, X) = P(U\mid S=1, A, X)/P(U\mid S=0, A, X),   V(X)=     E\{  Y^{(0)} \mid S=1, A=1, X\} -  E\{  Y^{(0)} \mid S=1, A=0, X\} 
- [E\{  Y^{(0)} \mid S=0, A=1, X\} -  E\{  Y^{(0)} \mid S=0, A=0, X\} ]$, and $T(X)= E \{ Y^{(1)} - Y^{(0)} \mid S=0, A=1, X\}$. From Assumption \ref{assump: S3}, we know 
\begin{align*}
	T(X) \in [- \Lambda, \Lambda]. 
\end{align*}
From the Bayes formula, we have $r(A, U, X) =\text{OR} \{\pi_S (U, A, X) , \pi_S (A, X)  \}  $, and thus $\Gamma^{-1} \leq r(A, U, X)\leq \Gamma $ under Assumption \ref{assump: S4}. Hence 
we show in Section \ref{sec: proof} that 
\begin{align}\label{eq: bound in marginal}
	\begin{split}
		& E[ Y \mid S=0, A=1, X]  (\Gamma^{-1} - 1) - E[ Y \mid S=0, A=0, X]  (\Gamma - 1) \leq \\
		& \qquad  V(X)  \leq  E[ Y \mid S=0, A=1, X]  (\Gamma - 1) - E[ Y \mid S=0, A=0, X]  (\Gamma^{-1} - 1) .
	\end{split}
\end{align}

Therefore, with $\theta_0 $ written as $ \theta_0 = E[ \Delta_1(X) - \Delta_0 (X) \mid S=1, A=1] +  E[ T(X) - V(X) \mid S=1, A=1]$, we can construct the bounds for $\theta_0$ as $[\theta_L, \theta_U]$, where 
$\theta_U= E[ \Delta_1(X) - \Delta_0 (X) \mid S=1, A=1] + \Lambda  -E[ E\{ Y \mid S=0, A=1, X\}\mid S=1, A=1]  (\Gamma^{-1} - 1) + E[E\{ Y \mid S=0, A=0, X\}\mid S=1, A=1]  (\Gamma - 1)  $ 
and 
$\theta_L = E[ \Delta_1(X) - \Delta_0 (X) \mid S=1, A=1] - \Lambda  -E[ E\{ Y \mid S=0, A=1, X\}\mid S=1, A=1]  (\Gamma - 1) + E[E\{ Y \mid S=0, A=0, X\}\mid S=1, A=1]  (\Gamma^{-1} - 1)  $. In correspondence to the three estimation approaches, the bounds in the sensitivity analysis can also be estimated in three ways. 

First, using the IPW estimator, we have 
\begin{align*}
	\hat\theta_{U, {\rm ipw}} &= \hat\theta_{\rm ipw} + \Lambda -(\Gamma^{-1}-1) \frac{1}{n_{11}} \sum_{i=1}^n \frac{ \pi_S(X_i; \hat\psi) \pi_A(X_i, 1; \hat\alpha ) }{(1- \pi_S(X_i; \hat\psi ) ) \pi_A (X_i, 0; \hat\alpha ) }(1- S_i ) A_i Y_i    \\
	&\quad + (\Gamma-1) \frac{1}{n_{11}} \sum_{i=1}^n  \frac{ \pi_S(X_i; \hat\psi) \pi_A(X_i, 1; \hat\alpha ) }{(1- \pi_S(X_i; \hat\psi ) ) (1-\pi_A (X_i, 0; \hat\alpha ) ) } (1- S_i ) (1-A_i) Y_i ,\\
	\hat\theta_{L, {\rm ipw}}  &= \hat\theta_{\rm ipw} - \Lambda -(\Gamma-1) \frac{1}{n_{11}} \sum_{i=1}^n \frac{ \pi_S(X_i; \hat\psi) \pi_A(X_i, 1; \hat\alpha ) }{(1- \pi_S(X_i; \hat\psi ) ) \pi_A (X_i, 0; \hat\alpha ) }(1- S_i ) A_i Y_i    \\
	&\quad + (\Gamma^{-1}-1) \frac{1}{n_{11}} \sum_{i=1}^n  \frac{ \pi_S(X_i; \hat\psi) \pi_A(X_i, 1; \hat\alpha ) }{(1- \pi_S(X_i; \hat\psi ) ) (1-\pi_A (X_i, 0; \hat\alpha ) ) } (1- S_i ) (1-A_i) Y_i. 
\end{align*}
Second, using the regression-based estimator, we have 
\begin{align*}
	\hat\theta_{U, {\rm reg}} &= \hat\theta_{\rm reg} + \Lambda -(\Gamma^{-1}-1) \frac{1}{n_{11}} \sum_{i: S_i = A_i = 1} \mu_Y(0, 1, X_i; \hat\beta) \\
	&\qquad + (\Gamma -1) \frac{1}{n_{11}} \sum_{i: S_i = A_i = 1} \mu_Y(0, 0, X_i; \hat\beta)  ,\\
	\hat\theta_{L, {\rm reg}} &= \hat\theta_{\rm reg} - \Lambda -(\Gamma-1) \frac{1}{n_{11}} \sum_{i: S_i = A_i = 1} \mu_Y(0, 1, X_i; \hat\beta) \\
	&\qquad + (\Gamma^{-1}-1) \frac{1}{n_{11}} \sum_{i: S_i = A_i = 1} \mu_Y(0, 0, X_i; \hat\beta) . 
\end{align*}
Finally, using the doubly-robust estimator, we have 
\begin{align*}
	\hat\theta_{U, {\rm dr}}  & = \hat\theta_{\rm dr} + \Lambda \\
	& -(\Gamma^{-1}-1) \frac{1}{n_{11}}   \sum_{i: S_i = A_i = 1}  \mu_Y(0, 1, X_i; \hat\beta)    \\
	&- (\Gamma^{-1}-1) \frac{1}{n_{11}}  \sum_{i=1}^n  \frac{ \pi_S(X_i; \hat\psi) \pi_A(X_i, 1; \hat\alpha ) }{(1- \pi_S(X_i; \hat\psi ) ) \pi_A (X_i, 0; \hat\alpha ) }(1- S_i ) A_i \{ Y_i  - \mu_Y(0, 1, X_i; \hat\beta)\}    \\
	& + (\Gamma -1) \frac{1}{n_{11}} \sum_{i: S_i = A_i = 1} \mu_Y(0, 0, X_i; \hat\beta) \\
	&+ (\Gamma -1) \frac{1}{n_{11}} \sum_{i=1}^n \frac{ \pi_S(X_i; \hat\psi) \pi_A(X_i, 1; \hat\alpha ) }{(1- \pi_S(X_i; \hat\psi ) ) (1-\pi_A (X_i, 0; \hat\alpha ) ) } (1- S_i ) (1-A_i) \{ Y_i  - \mu_Y(0,0, X_i; \hat\beta) \} \\
	\hat\theta_{L, {\rm dr}} &= \hat\theta_{\rm dr} - \Lambda \\
	& -(\Gamma-1) \frac{1}{n_{11}}   \sum_{i: S_i = A_i = 1}  \mu_Y(0, 1, X_i; \hat\beta)    \\
	&- (\Gamma-1) \frac{1}{n_{11}}  \sum_{i=1}^n  \frac{ \pi_S(X_i; \hat\psi) \pi_A(X_i, 1; \hat\alpha ) }{(1- \pi_S(X_i; \hat\psi ) ) \pi_A (X_i, 0; \hat\alpha ) }(1- S_i ) A_i \{ Y_i  - \mu_Y(0, 1, X_i; \hat\beta)\}    \\
	& + (\Gamma^{-1} -1) \frac{1}{n_{11}} \sum_{i: S_i = A_i = 1} \mu_Y(0, 0, X_i; \hat\beta) \\
	&+ (\Gamma^{-1} -1) \frac{1}{n_{11}} \sum_{i=1}^n \frac{ \pi_S(X_i; \hat\psi) \pi_A(X_i, 1; \hat\alpha ) }{(1- \pi_S(X_i; \hat\psi ) ) (1-\pi_A (X_i, 0; \hat\alpha ) ) } (1- S_i ) (1-A_i) \{ Y_i  - \mu_Y(0,0, X_i; \hat\beta) \}.
\end{align*}
Under the sensitivity model in Assumptions \ref{assump: S3}-\ref{assump: S4}, and $Y_i\geq 0$ and for fixed $\Gamma $ and $\Lambda$, the confidence interval for $\theta_0$ is 
\begin{align}
	[\hat\theta_{L, *} - z_{\alpha/2} \hat \sigma_{L, *}, \hat\theta_{U, *} + z_{\alpha/2} \hat \sigma_{U, *}],   \label{eq: SA CI 2}
\end{align}
where $z_{\alpha/2}$ is the $\alpha/2$-upper quantile of the standard normal distribution,  and $\hat\sigma_{L, *}, \hat\sigma_{U, *}$ are respectively the standard error of $\hat\theta_{L, *}, \hat\theta_{U, *}$ 
for $*\in \{\rm ipw, reg, dr\}$.

\subsection{EITC application: discussion of Assumption \ref{assump: parallel} and sensitivity analysis}

For the EITC application in Section 4, as discussed in the Supplement \S1.1, a sufficient condition for Assumption \ref{assump: parallel} is that (i) within each socioeconomic cell defined above, the smoking and drinking rates are similar for single mothers in the placebo sample and the primary sample; and (ii) the effect of smoking and drinking on birth weight is not modified by the mother's education.  Assumption \ref{assump: placebo samples} is likely to hold but there is a concern that Assumption \ref{assump: parallel} may be violated. A sensitivity analysis based on the regression-based estimator finds that the confidence interval in \eqref{eq: SA CI 2} becomes [-1.208, 0.069] with $\Lambda=0$ and $\Gamma=1.01$. This means that the observed treatment effect becomes insignificant when we allow for a small violation of Assumption \ref{assump: parallel}. Therefore, the conclusion that EITC reduces risk of low birth weight is sensitive to potential violations of Assumption \ref{assump: parallel}.


\subsection{Stabilized estimators}
The idea of stabilization is to normalize the weights to sum to one \citep{robins2007comment}.  The stabilized estimators are asymptotically equivalent to their unstabilized counterparts but they have better finite-sample performance. 

The stabilized IPW is 
\begin{align*}
	&\hat\theta_{\rm sipw}  \\
	&= \frac{1}{n_{11}}\sum_{i=1}^n S_i A_i Y_i \\
	&- \left\{   \sum_{i=1}^n  \frac{S_i (1-A_i) \pi_A(X_i, 1;\hat\alpha)}{1-\pi_A(X_i, 1;\hat\alpha)} \right\}^{-1} \sum_{i=1}^n \frac{S_i (1-A_i)\pi_A(X_i, 1;\hat\alpha)}{1-\pi_A(X_i, 1;\hat\alpha)} Y_i \\
	&- \left\{  \sum_{i=1}^n  \frac{(1-S_i) A_i\pi_S(X_i; \hat\psi)}{1-\pi_S(X_i; \hat\psi)}\frac{\pi_A(X_i, 1;\hat\alpha)}{\pi_A(X_i, 0;\hat\alpha)}  \right\}^{-1}  \sum_{i=1}^n \frac{(1-S_i) A_i \pi_S(X_i; \hat\psi)}{1-\pi_S(X_i; \hat\psi)}\frac{\pi_A(X_i, 1;\hat\alpha)}{\pi_A(X_i, 0;\hat\alpha)} Y_i\\
	&+\left\{  \sum_{i=1}^n  \frac{(1-S_i)(1-A_i)\pi_S(X_i; \hat\psi) }{1- \pi_S(X_i; \hat\psi)} \frac{ \pi_A(X_i, 1; \hat\alpha) }{ 1-  \pi_A (X_i, 0; \hat\alpha)} \right\}^{-1}
	\sum_{i=1}^n \frac{(1-S_i)(1-A_i) \pi_S(X_i; \hat\psi) }{1- \pi_S(X_i; \hat\psi)} \frac{ \pi_A(X_i, 1; \hat\alpha) }{ 1-  \pi_A (X_i, 0; \hat\alpha)} Y_i .
\end{align*}

\section{Technical Proofs}\label{sec: proof}
\subsection{Proof of Proposition \ref{prop: identification}}
The part (a) is directly from the assumptions. We only prove part (b).	First note that 
\begin{align*}
	&  E\left[ \frac{S - \pi_S(X) }{\pi_S(X)  \{ 1- \pi_S(X)\}} \frac{A - \pi_A(X,S)}{\pi_A(X,S) \{1 - \pi_A(X,S)\}}  Y \mid X\right] \\
	&=  E\left[ \frac{1 }{\pi_S(X) } \frac{A - \pi_A(X,1)}{\pi_A(X,1) \{1 - \pi_A(X,1)\}}  Y \mid X, S=1\right] P (S=1\mid X)  \\
	&\qquad - E\left[ \frac{ 1 }{ \{ 1- \pi_S(X)\}} \frac{A - \pi_A(X,0)}{\pi_A(X,0) \{1 - \pi_A(X,0)\}}  Y \mid X, S=0 \right] P (S=0\mid X)  \\
	&=  E\left[   \frac{A - \pi_A(X,1)}{\pi_A(X,1) \{1 - \pi_A(X,1)\}}  Y \mid X, S=1\right]    - E\left[   \frac{A - \pi_A(X,0)}{\pi_A(X,0) \{1 - \pi_A(X,0)\}}  Y \mid X, S=0 \right]   \\
	&= E\left[   \frac{A - \pi_A(X,1)}{\pi_A(X,1) \{1 - \pi_A(X,1)\}}  Y \mid X, S=1, A=1\right] P(A=1\mid X, S=1)  \\
	&\qquad + E\left[   \frac{A - \pi_A(X,1)}{\pi_A(X,1) \{1 - \pi_A(X,1)\}}  Y \mid X, S=1, A=0 \right] P(A=0 \mid X, S=1)    \\
	&\qquad - E\left[   \frac{A - \pi_A(X,0)}{\pi_A(X,0) \{1 - \pi_A(X,0)\}}  Y \mid X, S=0 , A=1 \right] P(A=1\mid X, S=0)   \\
	&\qquad - E\left[   \frac{A - \pi_A(X,0)}{\pi_A(X,0) \{1 - \pi_A(X,0)\}}  Y \mid X, S=0 , A=0  \right] P(A=0\mid X, S=0)   \\
	&=  E\left[    Y \mid X, S=1, A=1\right] - E\left[    Y \mid X, S=1, A=0 \right] \\
	&\qquad - E\left[    Y \mid X, S=0, A=1\right] +	 E\left[    Y \mid X, S=0, A=0 \right]  \\
	&= E\{ Y(1) - Y(0) \mid S=1, A=1, X \} . 
\end{align*}
Then, 
\begin{align*}
	\theta_0 &=  E \left(  E\left[ \frac{S - \pi_S(X) }{\pi_S(X)  \{ 1- \pi_S(X)\}} \frac{A - \pi_A(X,S)}{\pi_A(X,S) \{1 - \pi_A(X,S)\}}  Y \mid X\right]  \frac{ P(A= 1\mid S=1, X) P(S=1\mid X) }{P(S=1, A=1)} \right) \\
	&=  \frac{1}{P(S=1, A=1)} E\left[ \frac{S - \pi_S(X) }{  1- \pi_S(X)  } \frac{  \pi_A(X, 1)  \{ A - \pi_A(X,S)\} }{\pi_A(X,S) \{1 - \pi_A(X,S)\}}  Y  \right] . 
\end{align*}

\subsection{Proof of Theorem \ref{theo: EIF}}
In this section, we use subscripts to explicitly index quantities that depend on the distribution $P$, we use a zero subscript to denote a quantity evaluated at the true distribution $P= P_0$, we use a $\epsilon$ subscript to denote a quantity evaluated at the parametric submodel $P= P_\epsilon$. We will show that $\varphi (O; \theta_P, \eta_P)$ is equal to the efficient influence function by showing that it is the canonical gradient of the pathwise derivative of $\theta_P$, i.e., \citep{Newey1994} 
\begin{align}
	\frac{\partial \theta_\epsilon}{\partial \epsilon} \mid_{\epsilon = 0} = E_0 \{\text{EIF} (O; \theta_P, \eta_P)s_0(O) \} ,
\end{align}
where $\theta_\epsilon = \theta_{P_\epsilon}$, $s_\epsilon (O) = \partial \log dP_{\epsilon}(O)/\partial \epsilon$ denotes the parameter submodel score, which can be decomposed as $s_\epsilon (Y\mid S, A, X) + s_\epsilon (A\mid S,X) + s_\epsilon (S\mid X) + s_\epsilon (X)$ or $s_\epsilon (X\mid Y, S, A) + s_\epsilon (Y\mid S, A) +  s_\epsilon (S, A) $.

By the equality in Proposition \ref{prop: identification}(a), we have 
\begin{align}
	\begin{split}\label{eq: theta}
		\frac{\partial}{\partial \epsilon} \theta_\epsilon \mid_{\epsilon = 0}  & = \frac{\partial}{\partial \epsilon}  E_\epsilon(Y \mid S= 1, A=1)\mid_{\epsilon = 0} \\
		&\qquad -  \frac{\partial}{\partial \epsilon} E_\epsilon \left\{ E_\epsilon(Y \mid S= 1, A=0, X) \mid S=1, A=1\right\} \mid_{\epsilon = 0} \\
		&\qquad - \frac{\partial}{\partial \epsilon} E_\epsilon\left\{ E_\epsilon(Y\mid S=0, A=1, X)  \mid S=1, A=1\right\}\mid_{\epsilon = 0} \\
		&\qquad + \frac{\partial}{\partial \epsilon} E_\epsilon\left\{ E_\epsilon(Y\mid S=0, A=0, X)  \mid S=1, A=1\right\} \mid_{\epsilon = 0}\\
		&:= M_1 - M_2 - M_3 + M_4.
	\end{split}
\end{align}

In what follows, we use the following identities repeatedly: 
\begin{align}
	E_\epsilon \{b(O_1) s_\epsilon  ( O_2\mid O_1) \} = 0,  \label{eq: cond1} \\
	E_\epsilon \{b(O_1)  ( O_2 - E_\epsilon (O_2\mid O_1)) \} = 0,  \label{eq: cond2}
\end{align}
for any $b$ and any $(O_1, O_2)\subset O$. 

Consider each term $M_1, M_2, M_3, M_4$ separately: 
\begin{align*}
	M_1 &=   E_0 \{  Y s_0(Y\mid S, A)  \mid S=1, A=1\} \\
	&=  E_0 \left[ \frac{SA}{E_0(SA)} Y s_0(Y\mid S, A) \right] \\
	&= E_0 \left[ \frac{SA}{E_0(SA)} \{Y - E_0(Y\mid S, A)\} s_0(Y\mid S, A) \right] \quad \eqref{eq: cond1}\\
	&= E_0 \left[ \frac{SA}{E_0(SA)} \{Y - E_0(Y\mid S, A)\} \{ s_0(Y\mid S, A)+ s_0(S, A)\} \right] \quad \eqref{eq: cond2}\\
	&= E_0 \left[ \frac{SA}{E_0(SA)} \{Y - E_0(Y\mid S, A)\} \{ s_0(Y\mid S, A)+ s_0(S, A) + s_0(X\mid Y, S, A)\} \right] \quad \eqref{eq: cond1}\\
	&= E_0 \left[ \frac{SA}{E_0(SA)} \{Y - E_0(Y\mid S, A)\}  s_0(O) \right] \\
	M_2 & = E_0 \left\{ E_0 (Y s_0(Y\mid S, A, X) \mid S= 1, A=0, X) \mid S=1, A=1\right\} \\
	&\quad +  E_0 \left\{ E_0 (Y  \mid S= 1, A=0, X) s_0 (X\mid S, A) \mid S=1, A=1\right\} \\
	&= E_0 \left\{  \frac{SA}{E_0(SA)}  E_0 \left( \frac{S (1-A)}{P_0(S=1, A=0\mid X)} Y s_0(Y\mid S, A, X)\mid  X\right) \right\} \\
	&\quad +  E_0 \left\{  \frac{SA}{E_0(SA)}  E_0 (Y  \mid S= 1, A=0, X) s_0 (X\mid S, A) \right\} \\
	&= E_0 \left\{  \frac{E_0(SA\mid X)}{E_0(SA)} \frac{S (1-A)}{E_0 (S(1-A)\mid X)} Y s_0(Y\mid S, A, X) \right\} \\
	&\quad +  E_0 \left\{  \frac{SA}{E_0(SA)}  E_0 (Y  \mid S= 1, A=0, X) s_0 (X\mid S, A) \right\} \\
	&= E_0 \left\{  \frac{E_0 (SA\mid X) }{E_0(SA)} \frac{S (1-A)}{E_0 (S(1-A)\mid X)} Y s_0(Y\mid S, A, X) \right\} \\
	&\quad +  E_0 \left\{  \frac{SA}{E_0(SA)}  E_0 (Y  \mid S= 1, A=0, X) s_0 (X\mid S, A) \right\}\\
	&= E_0 \left\{  \frac{E_0(SA\mid X) }{E_0(SA)} \frac{S (1-A)}{E_0(S(1-A)\mid X)} \{Y -\mu_{Y0} (S, A, X) \} s_0(Y\mid S, A, X) \right\} \quad \eqref{eq: cond1}\\
	&\quad +  E_0 \left\{  \frac{SA}{E_0(SA)}  \{\mu_{Y0} (1, 0, X) -   E_0(\mu_{Y0} (1, 0, X)\mid S, A)\} s_0 (X\mid S, A) \right\}\quad \eqref{eq: cond1}\\
	&= E_0 \left\{  \frac{E_0(SA\mid X) }{E_0(SA)} \frac{S (1-A)}{E_0(S(1-A)\mid X)} \{Y - \mu_{Y0} (S, A, X)\} \{ s_0(Y\mid S, A, X) + s_0 (S, A, X)\} \right\} \quad \eqref{eq: cond2}\\
	&\quad +  E_0 \left\{  \frac{SA}{E_0(SA)}  \{\mu_{Y0} (1, 0, X) - E_0(\mu_{Y0} (1, 0, X)\mid S, A)\} \{ s_0 (X\mid S, A) +s_0(S, A)\}\right\}\quad \eqref{eq: cond2}\\
	&= E_0 \left\{  \frac{E_0(SA\mid X) }{E_0(SA)} \frac{S (1-A)}{E_0(S(1-A)\mid X)} \{Y - \mu_{Y0} (S, A, X)\} s_0(O) \right\} \\
	&\quad +  E_0 \left\{  \frac{SA}{E_0(SA)}  \{\mu_{Y0} (1, 0, X) - E_0(\mu_{Y0} (1, 0, X)\mid S, A)\} S_0(O) \right\}\quad \eqref{eq: cond1}\\
	M_3& =     E_0 \left\{  \frac{E_0(SA\mid X)}{E_0(SA)} \frac{(1-S) A }{E( (1-S) A\mid X)} Y s_0(Y\mid S, A, X) \right\} \\
	&\quad +  E_0 \left\{  \frac{SA}{E_0(SA)}  E_0 (Y  \mid S= 0, A=1, X) s_0 (X\mid S, A) \right\} \\
	&= E_0 \left\{  \frac{E_0(SA\mid X) }{E_0(SA)} \frac{(1-S) A}{E_0((1-S) A\mid X)} \{Y - \mu_{Y0} (S, A, X)\} s_0(O) \right\} \\
	&\quad +  E_0 \left\{  \frac{SA}{E_0(SA)}  \{\mu_{Y0} (0, 1, X) - E_0(\mu_{Y0} (0, 1, X)\mid S, A)\} s_0(O) \right\}\\
	M_4 &= E_0 \left\{  \frac{E_0(SA\mid X)}{E_0(SA)} \frac{(1-S) (1-A)}{E_0((1-S) (1-A)\mid X)} Y s_0(Y\mid S, A, X) \right\} \\
	&\quad +  E_0 \left\{  \frac{SA}{E_0(SA)}  E_0 (Y  \mid S= 0, A=0, X) s_0 (X\mid S, A) \right\} \\
	&= E_0 \left\{  \frac{E_0(SA\mid X) }{E_0(SA)} \frac{(1-S) (1-A)}{E_0((1-S) (1-A)\mid X)} \{Y - \mu_{Y0} (S, A, X)\} s_0(O) \right\} \\
	&\quad +  E_0 \left\{  \frac{SA}{E_0(SA)}  \{\mu_{Y0} (0, 0, X) - E_0(\mu_{Y0} (0, 0, X)\mid S, A)\} s_0(O) \right\}. 
\end{align*}
Combining the above derivations, we have 
\begin{align}
	\frac{\partial \theta_\epsilon}{\partial \epsilon} \mid_{\epsilon = 0} = E_0 \{\text{EIF} (O; \theta_P, \eta_P)s_0(O) \}, 
\end{align}
with $\text{EIF} (O; \theta_P, \eta_P)$ being  the efficient influence function.

\subsection{Proof of double robustness}
Let $\bar\eta = (\bar\mu_Y, \bar\pi_S, \bar\pi_A) $ be the limit of $\hat\eta$. Here we will show that $E[ \text{EIF} (O, \theta_0; \bar\eta)] = 0$ as long as either $\bar \mu_Y = \mu_{Y0}$ or $(\bar\pi_S,  \bar\pi_A)= (\pi_{S0}, \pi_{A0}) $. In this section, expectations $E= E_0$ are evaluated under $P_0$, but we drop the subscript for notational convenience.  

When $\mu_Y$ is correctly specified, we have $\bar\mu_Y = \mu_{Y0}$. Then, by iterative expectation, 
\begin{align*}
	E[ \text{EIF} (O, \theta_0; \bar\eta)] = 	& E \left[ \frac{SA}{E(SA)} \{ Y -  \mu_{Y0}(1,0,X) -  \mu_{Y0}(0,1,X) + \mu_{Y0}(0,0,X) - \theta_0\}  \right.  \\
	& \quad - \frac{S (1- A) \frac{\bar \pi_A (X, 1)}{1- \bar \pi_A(X,1)}}{E\left[ S (1- A) \frac{\bar \pi_A (X, 1)}{1- \bar \pi_A(X,1)} \right]} \{ Y- \mu_{Y0}(1, 0, X)\}  \\
	& \quad - \frac{ (1-S) A \frac{\bar \pi_A (X, 1)}{\bar \pi_A(X,0)}\frac{\bar \pi_S(X)}{1-\bar \pi_S(X)} }{E\left[ (1-S) A \frac{\bar \pi_A (X, 1)}{\bar  \pi_A(X,0)}\frac{\bar \pi_S(X)}{1-\bar \pi_S(X)} \right]} \{ Y- \mu_{Y0}(0, 1, X)\} \\
	& \left.\quad + \frac{ (1-S) (1-A) \frac{\bar \pi_A (X, 1)}{1- \bar \pi_A(X,0)}\frac{\bar \pi_S(X)}{1-\bar \pi_S(X)} }{E\left[ (1-S) (1-A) \frac{\bar \pi_A (X, 1)}{ 1- \bar \pi_A(X,0)}\frac{\bar \pi_S(X)}{1-\bar \pi_S(X)} \right]} \{ Y- \mu_{Y0}(0, 0, X)\} \right]\\
	&= E \left[  Y -  \mu_{Y0}(1,0,X) -  \mu_{Y0}(0,1,X) + \mu_{Y0}(0,0,X) - \theta_0 \mid S=1, A=1 \right]  \\
	&= 0 .
\end{align*}

When $\pi_A, \pi_S$ are correctly specified, we have $\bar\pi_S = \pi_{S0}$ and $\bar\pi_A = \pi_{A0}$.  Then, by iterative expectation, 
\begin{align*}
	E[ \text{EIF} (O, \theta_0; \bar\eta)] = & E \left[ \frac{SA}{E(SA)} \{ Y -  \bar \mu_Y(1,0,X) -  \bar\mu_Y(0,1,X) +\bar \mu_Y(0,0,X) - \theta_0\}   \right. \\
	& \quad - \frac{S (1- A) \frac{\pi_{A0} (X, 1)}{1- \pi_{A0}(X,1)}}{E\left[ S (1- A) \frac{\pi_{A0} (X, 1)}{1- \pi_{A0}(X,1)} \right]} \{ Y- \bar\mu_Y(1, 0, X)\}  \\
	& \quad - \frac{ (1-S) A \frac{\pi_{A0} (X, 1)}{\pi_{A0}(X,0)}\frac{\pi_{S0}(X)}{1-\pi_{S0}(X)} }{E\left[ (1-S) A \frac{\pi_{A0} (X, 1)}{ \pi_{A0}(X,0)}\frac{\pi_{S0}(X)}{1-\pi_{S0}(X)} \right]} \{ Y- \bar\mu_Y(0, 1, X)\} \\
	& \quad \left. + \frac{ (1-S) (1-A) \frac{\pi_{A0} (X, 1)}{1- \pi_{A0}(X,0)}\frac{\pi_{S0}(X)}{1-\pi_{S0}(X)} }{E\left[ (1-S) (1-A) \frac{\pi_{A0} (X, 1)}{ 1- \pi_{A0}(X,0)}\frac{\pi_{S0}(X)}{1-\pi_{S0}(X)} \right]} \{ Y- \bar\mu_Y(0, 0, X)\} \right]\\
	&= E[ Y\mid S=1, A=1] - E\left\{ \bar \mu_Y (1, 0, X) \mid S=1, A=1\right\}\\
	&\quad - E\left\{ \bar \mu_Y (0, 1, X) \mid S=1, A=1\right\} + E\left\{ \bar \mu_Y (0, 0, X) \mid S=1, A=1\right\} - \theta_0 \\
	&\quad - E\{ E(Y\mid S=1, A=0, X) - \bar\mu_Y(1, 0, X)\mid S=1, A=1\}   \\
	&\quad - E\{ E(Y\mid S=0, A=1, X) - \bar\mu_Y(0, 1, X)\mid S=1, A=1\}   \\
	&\quad + E\{ E(Y\mid S=0, A=0, X) - \bar\mu_Y(0, 0, X)\mid S=1, A=1\}   \\
	&= 0.
\end{align*}

\subsection{Proof of Theorem \ref{theo: asymptotics}}
In what follows, we will use $P\{ f(O)\} = \int f(O) dP$ to denote expectation treating the function $f$ as fixed; thus $P\{ f(O)\}$ is random when $f$ is random, and is different from the fixed quantity $E\{ f(O)\}$ which averages over randomness in both $f$ and $O$. 

Since $\hat \theta$ is a $Z$-estimator, using Theorem 5.31 of \cite{vanderVaart:2000book}, we have that
\begin{align*}
	\sqrt{n} (\hat \theta - \theta_0) =    & \sqrt{n} P\{ \text{EIF}(O; \theta_0, \hat\eta)\} +  n^{-1/2} \sum_{i=1}^n [ \text{EIF} (O_i; \theta_0, \bar\eta) - E\{ \text{EIF} (O_i; \theta_0, \bar\eta)\} ] \\
	&+ o_p\left(1+ \sqrt{n} P\{ \text{EIF} (O; \theta_0, \hat\eta) \} \right).
\end{align*}

Using standard central limit theorem, the second term is asymptotically normal, and is $O_p(1)$. Hence, the consistency and rate of convergence of $\hat\theta$ depend on the property of the first term. We analyze $\sqrt{n} P\{ \text{EIF}(O; \theta_0, \hat\eta)\}$ in the following. 

Define $\eta= (\mu_Y,  \pi_A, \pi_S,  \lambda)$, $\hat \eta= (\hat\mu_Y,  \hat\pi_A, \hat\pi_S,  \hat\lambda)$,  $\tilde \eta= (\hat\mu_Y,  \hat\pi_A, \hat\pi_S, \lambda_0)$, and  $ \eta_0= (\pi_{S0}, \pi_{A0}, \mu_{Y0}, \lambda_0)$. From
$$
\theta_0 = P\left[ \frac{SA}{E(SA)} \{\mu_{Y0}(1,1,X) - \mu_{Y0}(1,0, X) - \mu_{Y0}(0,1,X) + \mu_{Y0}(0,0,X) \}\right],
$$
we can write $P\{ \text{EIF}(O; \theta_0, \tilde\eta) \} $ as 
\begin{align*}
	&P\{ \text{EIF}(O; \theta_0, \tilde\eta) \} \\
	&= P\bigg[ \frac{SA}{ E(SA) } \{ Y -  \hat\mu_Y(1,0,X) -   \hat\mu_Y(0,1,X) +  \hat\mu_Y(0,0,X) - \theta_0\}   \\
	& \quad - \frac{S (1- A) \frac{ \hat\pi_A (X, 1)}{1-  \hat\pi_A(X,1)}}{E(SA)} \{ Y-  \hat\mu_Y(1, 0, X)\}   - \frac{ (1-S) A \frac{ \hat\pi_A (X, 1)}{ \hat\pi_A(X,0)}\frac{ \hat\pi_S(X)}{1- \hat\pi_S(X)} }{E(SA)} \{ Y-  \hat\mu_Y(0, 1, X)\} \\
	& \quad  + \frac{ (1-S) (1-A) \frac{ \hat\pi_A (X, 1)}{1-  \hat\pi_A(X,0)}\frac{ \hat\pi_S(X)}{1- \hat\pi_S(X)} }{E(SA)} \{ Y-  \hat\mu_Y(0, 0, X)\} \bigg]\\
	&= P\bigg[ \frac{SA}{ E(SA) } \{ -  \{\hat\mu_Y(1,0,X)-\mu_{Y0}(1,0,X)\}  -   \{\hat\mu_Y(0,1,X) -\mu_{Y0}(0,1,X)  \} \} \\
	&\quad + \frac{SA}{E(SA)}  \{\hat\mu_Y(0,0,X)- \mu_{Y0}(0,0,X) \} \}   \\
	& \quad - \frac{S (1- A) \frac{ \hat\pi_A (X, 1)}{1-  \hat\pi_A(X,1)}}{E(SA)} \{ \mu_{Y0} (1,0,X) -  \hat\mu_Y(1, 0, X)\}  \\
	&\quad - \frac{ (1-S) A \frac{ \hat\pi_A (X, 1)}{ \hat\pi_A(X,0)}\frac{ \hat\pi_S(X)}{1- \hat\pi_S(X)} }{E(SA)} \{ \mu_{Y0} (0,1,X)-  \hat\mu_Y(0, 1, X)\} \\
	& \quad  + \frac{ (1-S) (1-A) \frac{ \hat\pi_A (X, 1)}{1-  \hat\pi_A(X,0)}\frac{ \hat\pi_S(X)}{1- \hat\pi_S(X)} }{E(SA)} \{ \mu_{Y0} (0,0,X)-  \hat\mu_Y(0, 0, X)\} \bigg]\\
	&=  P\bigg[ \bigg\{ \frac{SA}{E(SA)} - \frac{S (1- A) \frac{ \hat\pi_A (X, 1)}{1-  \hat\pi_A(X,1)}}{E(SA)} \bigg\}\{ \mu_{Y0} (1,0,X) -  \hat\mu_Y(1, 0, X)\}  \bigg]\\
	&\quad+ P\bigg[ \bigg\{ \frac{SA}{E(SA)} - \frac{ (1-S) A \frac{ \hat\pi_A (X, 1)}{ \hat\pi_A(X,0)}\frac{ \hat\pi_S(X)}{1- \hat\pi_S(X)} }{E(SA)} \bigg\} \{ \mu_{Y0} (0,1,X)-  \hat\mu_Y(0, 1, X)\} \bigg] \\
	& \quad  + P\bigg[ \bigg\{ - \frac{SA}{E(SA)}+ \frac{ (1-S) (1-A) \frac{ \hat\pi_A (X, 1)}{1-  \hat\pi_A(X,0)}\frac{ \hat\pi_S(X)}{1- \hat\pi_S(X)} }{E(SA)} \bigg\} \{ \mu_{Y0} (0,0,X)-  \hat\mu_Y(0, 0, X)\} \bigg]\\
	&:= T_1 + T_2 + T_3  .
\end{align*}

From the facts that 
\begin{align*}
	&\| \hat \pi_A (X, S) - \pi_{A0} (X, S)\|_2^2 \\
	&=   \|  \{\hat \pi_A (X,1) - \pi_{A0} (X,1) \} \pi_{S0}(X)^{1/2}\|_2^2  +   \| \{\hat \pi_A (X,0) - \pi_{A0} (X,0)\} \{ 1-\pi_{S0}(X) \}^{1/2}\|_2^2 \\
	&\| \hat \mu_Y (S,A, X) - \mu_{Y0} (S,A, X)\|_2^2 \\
	&=  \| \{ \hat \mu_Y (1,1, X) - \mu_{Y0} (1,1, X)\} \pi_{S0}(X)^{1/2}\pi_{A0}(X,1)^{1/2} \|_2^2 \\
	&\qquad +\| \{ \hat \mu_Y (1,0, X) - \mu_{Y0} (1,0, X)\} \pi_{S0}(X)^{1/2}(1-\pi_{A0}(X,1))^{1/2} \|_2^2 \\
	&\qquad +\| \{ \hat \mu_Y (0,1, X) - \mu_{Y0} (0, 1, X)\} (1-\pi_{S0}(X))^{1/2}\pi_{A0}(X,1)^{1/2} \|_2^2 \\
	&\qquad +\| \{ \hat \mu_Y (0,0, X) - \mu_{Y0} (0, 0, X)\} (1-\pi_{S0}(X))^{1/2} (1-\pi_{A0}(X,1))^{1/2} \|_2^2, 
\end{align*}
we have that every term on the right-hand side of the equations are bounded by the term on the left-hand side, e.g.,  
\begin{align*}
	&  \|  \{\hat \pi_A (X,1) - \pi_{A0} (X,1) \} \pi_{S0}(X)^{1/2}\|_2  \leq\| \hat \pi_A (X,S) - \pi_{A0} (X,S)\|_2. 
\end{align*}

Hence, from Cauchy-Schwarz inequality, boundedness of $1/\hat\pi_A(X, 0), 1/\{1- \hat\pi_A(X, 0)\},1/\{1- \hat\pi_A(X, 1)\}, 1/\{ 1- \hat \pi_S(X)\}$, and  the triangle inequality, the first term satisfies
\begin{align*}
	T_1 &= P\bigg[  \frac{ \pi_{S0}(X)^{1/2} \{ \pi_{A0} (X,1)- \hat\pi_A(X, 1)\}}{E(SA)\{1- \hat\pi_A(X, 1)\}} \frac{\pi_{S0}(X)^{1/2}(1-\pi_{A0}(X, 1) )^{1/2}}{(1-\pi_{A0}(X, 1) )^{1/2} } \{ \mu_{Y0} (1,0,X) -  \hat\mu_Y(1, 0, X)\} \bigg] \\
	&\leq C \| \pi_{S0}(X)^{1/2} \{ \pi_{A0}(X,1) - \hat \pi_A(X,1)\}\|_2 \|\pi_{S0}(X)^{1/2}(1-\pi_{A0}(X, 1) )^{1/2}\{ \mu_{Y0} (1, 0, X) - \hat \mu_Y (1, 0, X)\}\|_2 \\
	&= O_p( \| \hat \pi_A (X,S) - \pi_{A0} (X,S)\|_2 \| \hat \mu_Y (S,A, X) - \mu_{Y0} (S,A, X)\|_2  ).
\end{align*}
Similarly, for the second and third terms, we have  
\begin{align*}
	T_2 & = P\bigg[ \frac{\pi_{A0}(X,1) \pi_{S0}(X) \hat \pi_A(X,0) (1-\hat\pi_S(X)) - \pi_{A0}(X,0) (1-\pi_{S0}(X)) \hat \pi_A(X,1) \hat\pi_S(X)}{E(SA) \hat \pi_A(X,0) (1-\hat\pi_S(X))} \cdot \\
	& \qquad \qquad\qquad\qquad\qquad\qquad\qquad\qquad\qquad \qquad\qquad\qquad   \{ \mu_{Y0} (0,1,X)-  \hat\mu_Y(0, 1, X)\} \bigg] \\
	&= O_p\left( \{ \| \hat \pi_S(X) - \pi_{S0}(X)\|_2 +   \| \hat \pi_A (X,S) - \pi_{A0} (X,S)\|_2 \} \| \hat\mu_Y(S, A, X) - \mu_{Y0} (S, A, X)\|_2 \right)\\
	T_3 &=- \frac{\pi_A(X,1) \pi_{S0} (X) (1- \hat \pi_A(X,0) ) (1-\hat\pi_S(X)) -(1- \pi_{A0}(X,0)) (1-\pi_{S0}(X)) \hat \pi_A(X,1) \hat\pi_S(X)}{E(SA)  (1- \hat\pi_A(X,0)) (1-\hat\pi_S(X))} \\
	&\qquad \qquad\qquad\qquad\qquad\qquad\qquad\qquad\qquad \qquad\qquad\qquad \{ \mu_{Y0}(0,0,X) - \hat\mu_Y(0,0,X)\}\bigg] \\
	& = O_p\left( \{ \| \hat \pi_S(X) - \pi_{S0}(X)\|_2 +   \| \hat \pi_A (X,S) - \pi_{A0} (X,S)\|_2 \} \| \hat\mu_Y(S, A, X) - \mu_{Y0} (S, A, X)\|_2 \right).
\end{align*}
Therefore, 
\[
P\{ \text{EIF}(O; \theta_0, \tilde\eta) \} = O_p\left( \{ \| \hat \pi_S - \pi_{S0}\|_2 +   \| \hat \pi_A  - \pi_{A0} \|_2 \} \| \hat\mu_Y - \mu_{Y0} \|_2 \right)
\]

Similarly, we can show that 
\begin{align*}
	&P\{ \varphi(O; \theta_0, \tilde\eta)\} - P\{ \varphi(O; \theta_0, \hat\eta)\}  = O_p\left\{  \left( \|\hat\pi_S - \pi_{S0} \|_2  + \| \hat\pi_A - \pi_{A0} \|_2 \right)\| \hat\mu_Y -\mu_{Y0} \|_2  |\hat \lambda - \lambda_0| \right\}.
\end{align*}

Again applying the triangle inequality and because $\hat\lambda- \lambda_0 = o_p(1)$,  we arrive at 
\[
P\{ \varphi(O; \theta_0, \hat\eta)\} = O_p\left\{  \big( \| \hat\pi_A - \pi_{A0} \|_2  +  \|\hat\pi_S - \pi_{S0} \|_2  \big) \| \hat\mu_Y -\mu_{Y0} \|_2 \right\}.
\]

\subsection{Proof of \eqref{eq: bound in marginal}}

Recall the definition $   V(X) =     E\{  Y^{(0)} \mid S=1, A=1, X\} -  E\{  Y^{(0)} \mid S=1, A=0, X\} - [E\{  Y^{(0)} \mid S=0, A=1, X\} -  E\{  Y^{(0)} \mid S=0, A=0, X\} ]$. Equation \eqref{eq: bound in marginal} follows from the result that 
\begin{align*}
	V(X) &=  E\left[ E\{Y \mid S=0, U,X \} \{ r(1, U, X) - 1\} \mid S=0, A=1, X \right]  \\
	&\quad  - E\left[ E\{Y\mid S=0, U,X \} \{ r(0, U, X) - 1\} \mid S=0, A=0, X \right] .
\end{align*}

\section{Variance estimator} \label{sec: variance}
\subsection{Constructing a consistent variance estimator}
A general recipe for constructing a consistent variance estimator under the M-estimation framework (see, e.g., tutorials by \citet{stefanski2002calculus} and \citet{saul2020calculus}) can be applied to our proposed estimators as follows. Let $\text{IF}(O; \theta, \eta)$ denote the efficient influence function corresponding to $\hat\theta_{\rm reg}$, $\hat\theta_{\rm ipw}$, or
$\hat\theta_{\rm dr}$. Nuisance parameters $\eta$ are estimated by $\hat\eta$ via solving their corresponding estimating equations $\sum_{i=1}^n\phi(O_i; \hat\eta) = 0$, and the target parameter $\theta$ is estimated by $\hat{\theta}$ via solving $\sum_{i=1}^n\text{IF}(O_i; \hat\theta, \hat\eta) = 0$. Write $\gamma = (\theta, \eta)$ for simplicity. To estimate $\gamma$, it suffices to stack estimating equations $\phi(O_i; \eta)$ and $\text{IF}(O_i; \theta, \eta)$, and find $\gamma = \hat\gamma$ that solves 
\begin{equation}\sum_{i = 1}^n \psi(O_i; \hat\gamma) =
	\begin{pmatrix}
		\sum_{i=1}^n\phi(O_i; \hat\eta) \\
		\sum_{i=1}^n\text{IF}(O_i; \hat\theta, \hat\eta)
	\end{pmatrix}=
	\begin{pmatrix}
		0 \\
		0
	\end{pmatrix}.
\end{equation}
Define $A(\gamma) = \mathbb{E}\{-\partial \psi(O; \gamma)/\partial \gamma\}$ and $B(\gamma) = \mathbb{E}\{\psi(O; \gamma)\psi(O; \gamma)^T\}$. A consistent estimator of the variance-covariance matrix of $\hat\gamma$ can then be obtained using 
\begin{equation}
	\hat V(\hat\gamma) = (1/n)\cdot A(\hat\gamma)^{-1}B(\hat\gamma)\{A(\hat\gamma)^{-1}\}^T.
\end{equation}
A consistent variance estimator of the target parameter $\theta$ is then obtained as the $(1,1)$-th element of $ \hat V(\hat\gamma)$.

\subsection{Implementation in \textsf{R} using the \textsf{geex} package}
The \textsf{R} package \textsf{geex} is a general purpose software to find roots and compute the empirical sandwich estimator for a set of user-supplied, unbiased estimating equations (\citealp{saul2020calculus}). Below, we provide an implementation of the sandwich variance estimator via the \textsf{geex} package.

\begin{lstlisting}
	###############################################
	# Estimating equation
	dr_est_func <- function(data, models){
		
		A = data$A
		S = data$S
		Y = data$Y
		
		Xs <- grab_design_matrix(data = data,
		rhs_formula = grab_fixed_formula(models$S_md))
		
		Xa = grab_design_matrix(data = data,
		rhs_formula = grab_fixed_formula(models$A_md))
		
		Xmu <- grab_design_matrix(data = data,
		rhs_formula = grab_fixed_formula(models$mu_md))
		
		Xa_S1 = Xa
		Xa_S1[, 'S'] = 1
		
		Xa_S0 = Xa
		Xa_S0[, 'S'] = 0
		
		Xmu_S1_A0 = Xmu
		Xmu_S1_A0[, 'S'] = 1
		Xmu_S1_A0[, 'A'] = 0
		Xmu_S1_A0[, 'S:A'] = 0
		
		Xmu_S0_A1 = Xmu
		Xmu_S0_A1[, 'S'] = 0
		Xmu_S0_A1[, 'A'] = 1
		Xmu_S0_A1[, 'S:A'] = 0
		
		Xmu_S0_A0 = Xmu
		Xmu_S0_A0[, 'S'] = 0
		Xmu_S0_A0[, 'A'] = 0
		Xmu_S0_A0[, 'S:A'] = 0
		
		S_md_pos = 1:length(coef(models$S_md))
		S_scores <- grab_psiFUN(models$S_md, data = data)
		
		A_md_pos = (length(coef(models$S_md))+1):(length(coef(models$S_md)) + 
		length(coef(models$A_md)))
		A_scores <- grab_psiFUN(models$A_md, data = data)
		
		mu_md_pos = (length(coef(models$S_md)) + length(coef(models$A_md)) + 1):
		(length(coef(models$S_md)) + length(coef(models$A_md)) +
		length(coef(models$mu_md)))
		mu_scores <- grab_psiFUN(models$mu_md, data = data)
		
		function(theta, n, n_11) {
			pi_s = plogis(Xs %*% theta[S_md_pos])
			pi_a_s_1 = plogis(Xa_S1 %*% theta[A_md_pos])
			pi_a_s_0 = plogis(Xa_S0 %*% theta[A_md_pos])
			
			mu_s_1_a_0 = Xmu_S1_A0 %*% theta[mu_md_pos]
			mu_s_0_a_1 = Xmu_S0_A1 %*% theta[mu_md_pos]
			mu_s_0_a_0 = Xmu_S0_A0 %*% theta[mu_md_pos]
			
			beta = S * A * (n/n_11) * (Y - mu_s_1_a_0 - mu_s_0_a_1 + 
			mu_s_0_a_0 - theta[length(theta)]) - 
			(n/n_11) * S * (1 - A) * (pi_a_s_1/(1-pi_a_s_1)) * 
			(Y - mu_s_1_a_0) - (n/n_11) * (1 - S) * A * 
			(pi_a_s_1/pi_a_s_0) * (pi_s/(1-pi_s)) * 
			(Y - mu_s_0_a_1) + (n/n_11) * (1 - S) * (1 - A) * 
			(pi_a_s_1/(1-pi_a_s_0)) * (pi_s/(1-pi_s)) * (Y - mu_s_0_a_0)
			
			c(S_scores(theta[S_md_pos]),
			A_scores(theta[A_md_pos]),
			mu_scores(theta[mu_md_pos]),
			beta)
		}
	}
	
	#########################################################
	# Solve the estimating equation and compute the empirical
	# sandwich estimator. The function returns the point 
	# estimator and the estimated standard error.
	
	estimate_aipw <- function(data, models){
		
		n_11 = length(which(data$S == 1 & data$A == 1))
		n = dim(data)[1]
		
		res = m_estimate(estFUN = dr_est_func,
		data = data,
		root_control = setup_root_control(start = 
		c(unlist(lapply(models, function(x) coef(x))), 0)),
		outer_args = list(models = models),
		inner_args = list(n = n, n_11 = n_11))
		nparam = length(roots(res))
		point_est = roots(res)[nparam]
		se = sqrt(vcov(res)[nparam, nparam])
		return(c(point_est, se))
	}
	
	#######################################################
	# Run the code
	res = estimate_aipw(data = data, 
	models = list(S_md = S_md,
	A_md = A_md,
	mu_md = mu_md))
	
\end{lstlisting}

\section{Additional simulation results}

\begin{figure}
	\centering
	\includegraphics[width=\textwidth]{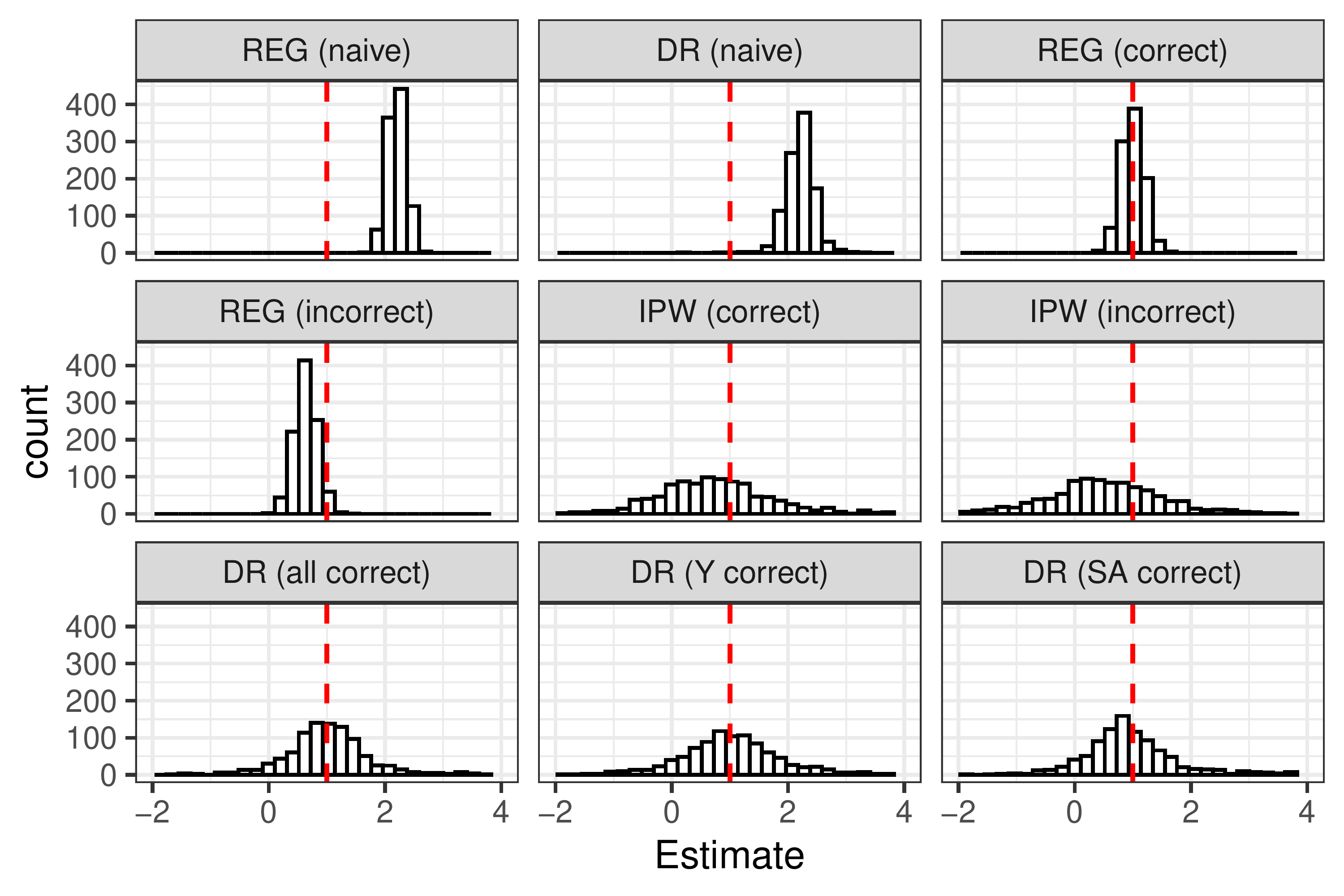}
	\caption{Sampling distribution of nine estimators under Scenario I.}
	\label{fig: sampling distribution}
\end{figure}

\begin{table}[ht]
	\centering
	\caption{Additional simulation results. Scenario IV: d = (d1), e = (e2), f = (f1). Scenario V: d = (d1), e = (e2), f = (f2). Scenario VI: d = (d2), e = (e1), f = (f2). $\theta_0 = 1$ in all three scenarios. We trimmed the 1\% tail of the most extreme values of each estimator when reporting the bias.}
	\label{tbl: simu n = 1000 supp 1}
	\resizebox{\textwidth}{!}{
		\begin{tabular}{lcccccccccc}
			\hline
			Estimator & \begin{tabular}{c}Model \\ Spec.\end{tabular} & Bias &\begin{tabular}{c}Median \\ Est. SE\end{tabular}& \begin{tabular}{c}Cov. \\ $95\%$ CI\end{tabular}& Bias &\begin{tabular}{c}Median \\ Est. SE\end{tabular}& \begin{tabular}{c}Cov. \\ $95\%$ CI\end{tabular}& Bias &\begin{tabular}{c}Median \\ Est. SE\end{tabular}& \begin{tabular}{c}Cov. \\ $95\%$ CI\end{tabular} \\ 
			\hline
			& &\multicolumn{3}{c}{Scenario IV} &\multicolumn{3}{c}{Scenario V}&\multicolumn{3}{c}{Scenario VI}\\
			$\hat\theta_{\rm reg, naive}$ & & 1.19 & 0.16 & 0\% & 1.20 & 0.16 & 0\% & 1.16 & 0.16 & 0\% \\ 
			$\hat\theta_{\rm dr, naive}$ & & 1.19 & 0.19 & 1.3\% & 1.20 & 0.19 &2.3\% & 1.60 & 0.17 & 1.2\% \\ 
			$\hat\theta_{\rm reg}$ & \begin{tabular}{c}$\mu_Y$ \\ correct\end{tabular} & 0.00 & 0.20 & 93.8\% & 0.00 & 0.20 & 95.5\% & 0.01 & 0.19 & 94.9\% \\ 
			$\hat\theta_{\rm reg}$ & \begin{tabular}{c}$\mu_Y$ \\ incorrect\end{tabular} & -0.29 & 0.20 & 66.5\% & -0.30 & 0.20 & 66.8\% & -0.35 & 0.20 & 57.4\% \\ 
			$\hat\theta_{\rm ipw}$ &\begin{tabular}{c}$(\pi_S, \pi_A)$ \\ correct\end{tabular} & -0.24 & 0.72 & 90.0\% & -0.19 & 0.70 & 91.2\% & -0.19 & 0.55 & 90.8\% \\ 
			$\hat\theta_{\rm ipw}$ &\begin{tabular}{c}$(\pi_S, \pi_A)$ \\ incorrect\end{tabular} & -0.60 & 0.76 & 84.8\% & -0.57 & 0.76 & 86.2\% & -0.43 & 0.56 & 84.9\% \\ 
			$\hat\theta_{\rm dr}$ & \begin{tabular}{c}All \\ correct\end{tabular} & -0.01 & 0.47 & 95.2\% & -0.02 & 0.47 & 93.6\% & -0.03 & 0.46 & 94.7\% \\ 
			$\hat\theta_{\rm dr}$ &  \begin{tabular}{c}$\mu_Y$ \\ correct\end{tabular} & -0.02 & 0.63 & 95.1\% & -0.01 & 0.62 & 94.6\% & 0.00 & 0.56 & 94.4\% \\ 
			$\hat\theta_{\rm dr}$ & \begin{tabular}{c}$(\pi_S, \pi_A)$ \\ correct\end{tabular} & -0.12 & 0.52 & 91.0\% & -0.08 & 0.49 & 91.8\% & -0.07 & 0.49 & 91.5\% \\ 
			\hline
	\end{tabular}}
\end{table}

\begin{table}[ht]
	\centering
	\caption{Additional simulation results. Scenario VII: d = (d2), e = (e2), f = (f1). Scenario VIII: d = (d2), e = (e2), f = (f2). $\theta_0 = 1$ in all three scenarios. We trimmed the 1\% tail of the most extreme values of each estimator when reporting the bias.}
	\label{tbl: simu n = 1000 supp 2}
	\resizebox{\textwidth}{!}{
		\begin{tabular}{lccccccc}
			\hline
			Estimator & \begin{tabular}{c}Model \\ Spec.\end{tabular} & Bias &\begin{tabular}{c}Median \\ Est. SE\end{tabular}& \begin{tabular}{c}Cov. \\ $95\%$ CI\end{tabular}& Bias &\begin{tabular}{c}Median \\ Est. SE\end{tabular}& \begin{tabular}{c}Cov. \\ $95\%$ CI\end{tabular} \\ 
			\hline
			& &\multicolumn{3}{c}{Scenario VII} &\multicolumn{3}{c}{Scenario VIII}\\
			$\hat\theta_{\rm reg, naive}$ & & 1.17 & 0.16 & 0\% & 1.17 & 0.17 & 0\% \\ 
			$\hat\theta_{\rm dr, naive}$ & & 1.20 & 0.16 & 0.8\% & 1.20 & 0.16 & 1.0\% \\ 
			$\hat\theta_{\rm reg}$ & \begin{tabular}{c}$\mu_Y$ \\ correct\end{tabular} & 0.00 & 0.20 & 94.5\% & 0.01 & 0.20 & 96.0\% \\ 
			$\hat\theta_{\rm reg}$ & \begin{tabular}{c}$\mu_Y$ \\ incorrect\end{tabular} & -0.30 & 0.20 & 69.2\% & -0.29 & 0.20 & 70.8\% \\ 
			$\hat\theta_{\rm ipw}$ &\begin{tabular}{c}$(\pi_S, \pi_A)$ \\ correct\end{tabular} & -0.15 & 0.63 & 91.4\% & -0.18 & 0.63 & 92.0\% \\ 
			$\hat\theta_{\rm ipw}$ &\begin{tabular}{c}$(\pi_S, \pi_A)$ \\ incorrect\end{tabular} & -0.51 & 0.66 & 87.0\% & -0.56 & 0.66 & 86.6\% \\ 
			$\hat\theta_{\rm dr}$ & \begin{tabular}{c}All \\ correct\end{tabular}  & 0.02 & 0.44 & 94.3\% & -0.05 & 0.45 & 93.1\% \\ 
			$\hat\theta_{\rm dr}$ &  \begin{tabular}{c}$\mu_Y$ \\ correct\end{tabular} & -0.08 & 0.55 & 94.7\% & -0.13 & 0.56 & 92.9\% \\ 
			$\hat\theta_{\rm dr}$ & \begin{tabular}{c}$(\pi_S, \pi_A)$ \\ correct\end{tabular} & -0.07 & 0.49 & 91.8\% & -0.14 & 0.50 & 90.4\% \\ 
			\hline
	\end{tabular}}
\end{table}

\end{document}